\documentclass[%
 aip,
 amsmath,amssymb,
 reprint,%
]{revtex4-1}

\usepackage{graphicx}
\usepackage{dcolumn}
\usepackage{bm}

\usepackage[utf8]{inputenc}
\usepackage[T1]{fontenc}
\usepackage{mathptmx}
\usepackage{etoolbox}
\usepackage{makecell}
\usepackage{array}
\usepackage{makecell}

\usepackage{soul}
\usepackage{color, xcolor}
\usepackage{CJKutf8}
\usepackage{hyperref}
\hypersetup{
    colorlinks = true,
    urlcolor    = red,
    linkcolor  = blue,
    citecolor  = blue,
    filecolor   = black,
}

\makeatletter
\def\@email#1#2{%
 \endgroup
 \patchcmd{\titleblock@produce}
  {\frontmatter@RRAPformat}
  {\frontmatter@RRAPformat{\produce@RRAP{*#1\href{mailto:#2}{#2}}}\frontmatter@RRAPformat}
  {}{}
}%
\makeatother
\begin{document}

\preprint{AIP/123-QED}

\title{Plasma kinetics: Discrete Boltzmann modeling and Richtmyer-Meshkov instability}
\author{Jiahui Song \begin{CJK*}{UTF8}{gbsn} (宋家辉) \end{CJK*}}
\affiliation{School of Aerospace Engineering, Beijing Institute of Technology, Beijing, 100081, P.R.China}
\affiliation{National Key Laboratory of Computational Physics, Institute of Applied Physics and Computational Mathematics,Beijing 100088, P.R.China}
\affiliation{National Key Laboratory of Shock Wave and Detonation Physics, Mianyang 621900, China}

\author{Aiguo Xu \begin{CJK*}{UTF8}{gbsn} (许爱国) \end{CJK*}}
 \altaffiliation{Corresponding author: Xu\_Aiguo@iapcm.ac.cn}
\affiliation{National Key Laboratory of Computational Physics, Institute of Applied Physics and Computational Mathematics,Beijing 100088, P.R.China}
\affiliation{HEDPS, Center for Applied Physics and Technology, and College of Engineering, Peking University, Beijing 100871,China}
\affiliation{State Key Laboratory of Explosion Science and Technology, Beijing Institute of Technology, Beijing 100081,China}

\author{Long Miao \begin{CJK*}{UTF8}{gbsn} (苗龙) \end{CJK*}}
\altaffiliation{Corresponding author: miaolong@bit.edu.cn}
\affiliation{School of Aerospace Engineering, Beijing Institute of Technology, Beijing, 100081, P.R.China}
\affiliation{Chongqing Innovation Center, Beijing Institute of Technology, Chongqing  404100, China}

\author{Feng Chen \begin{CJK*}{UTF8}{gbsn} (陈锋) \end{CJK*}}
\affiliation{School of Aeronautics, Shandong Jiaotong University, Jinan 250357, China}

\author{Zhipeng Liu \begin{CJK*}{UTF8}{gbsn} (刘枝朋) \end{CJK*}}
\affiliation{Department of Physics, School of Science, Tianjin Chengjian University, Tianjin, 300384, China}

\author{Lifeng Wang \begin{CJK*}{UTF8}{gbsn} (王立锋) \end{CJK*}}
\affiliation{National Key Laboratory of Computational Physics, Institute of Applied Physics and Computational Mathematics,Beijing 100088, P.R.China}
\affiliation{HEDPS, Center for Applied Physics and Technology, and College of Engineering, Peking University, Beijing 100871,China}

\author{Ningfei Wang \begin{CJK*}{UTF8}{gbsn} (王宁飞) \end{CJK*}}
\affiliation{School of Aerospace Engineering, Beijing Institute of Technology, Beijing, 100081, P.R.China}

\author{Xiao Hou \begin{CJK*}{UTF8}{gbsn} (侯晓) \end{CJK*}}
\affiliation{School of Aerospace Engineering, Beijing Institute of Technology, Beijing, 100081, P.R.China}

\date{\today}

\begin{abstract}
In this paper, a discrete Boltzmann model (DBM) for plasma kinetics is proposed and further used to investigate the non-equilibrium characteristics in Orszag-Tang (OT) vortex and Richtmyer-Meshkov instability (RMI) problems. 
The construction of DBM mainly considers two aspects. 
The first is to build a physical model with sufficient capability to capture underlying physics. 
The second is to devise schemes for extracting more valuable information from massive data. 
For the first aspect, the generated model is equivalent to a magnetohydrodynamic model, and a coarse-grained model for extracting the most relevant thermodynamic non-equilibrium (TNE) behaviors including the entropy production rate. 
For the second aspect, the DBM uses non-conserved kinetic moments of ($f-f^{eq}$) to describe the non-equilibrium states and behaviors of complex systems. 
It is found that: (i) For OT vortex, the entropy production rate and compression difficulty first increase then decrease with time.
(ii) For RMI with interface inversion and re-shock process, the influence of magnetic field on TNE effects shows stages: before the interface inversion, the TNE strength is enhanced by delaying the interface inversion; while after the interface inversion, the TNE strength is significantly reduced. Both the global average TNE strength and entropy production rate contributed by non-organized energy flux can be used as physical criteria to identify whether or not the magnetic field is sufficient to prevent the interface inversion. 
In general, this paper proposes a generalized physical modeling scheme that has the potential for investigating the kinetic physics in plasma.
\end{abstract}


\maketitle

\section{Introduction}
\label{Introduction}
As the fourth state of matter, plasma widely exists in nature and various industrial fields such as initial confinement fusion (ICF). \cite{betti2016Inertial,abu2022lawson} 
In ICF, the target pellet is driven by strong lasers or x-rays, which ionize the shell ablator material and form strong shock waves to compress the fuel centrally to the high temperature and density of the ignition state.
When the imposing shock wave passes through a perturbed interface, the perturbation amplitude will increase with time, resulting in the ablator material being mixed into the fuel plasmas and causing ignition failure. \cite{brouillette2002richtmyer}
This phenomenon was first investigated theoretically by Richtmyer\cite{Richtmyer1960} through linear stability analysis, then qualitatively verified by Meshkov \cite{meshkov1969instability} through shock tube experiments.
Now, it is generally referred to as the Richtmyer-Meshkov instability (RMI).
As a physical phenomenon that is bound to occur when certain conditions are met, RMI also plays an important role in astrophysics \cite{arnett2000role,sano2021laser}, shock wave physics, combustion \cite{khokhlov1999numerical} and other kinds of systems \cite{zhou20171,zhou20172,zhou2021rayleigh}. 

Due to the importance and complexity of RMI, extensive theoretical, experimental, and numerical studies have been conducted. \cite{zhai2011evolution,xu2016complex,lei2017experimental,zhou20171,zhou20172,wang2017theoretical,zhai2018review,chen2023numerical,
li2023richtmyer,wang2023high,huang2022richtmyer}
Among them, plasma RMI is one of the important research areas.
Previous theoretical and numerical studies of plasma RMI are mainly based on two kinds of models. 
The first kind is various magnetohydrodynamics (MHD) models that simultaneously couple Euler/Navier-Stokes (NS) equations with Maxwell's equations. 
Based on MHD models, the effects of applied magnetic fields and self-generated electromagnetic fields on the development of plasma RMI have been extensively studied. \cite{samtaney2003,wheatley2005,wheatley2005stability,wheatley2009richtmyer,
sano2013,wheatley2014transverse,mostert2015,bond2017richtmyer,zhang2020numerical,
qin2021richtmyer,li2022linear,bakhsh2022linear,zhang2023suppression}
Though various magnetic fluid models were developed, such as single-fluid MHD, two-fluid MHD, Hall MHD, ideal MHD, etc., the function of MHD models is still challenged in several aspects:

(i) The MHD assumes that the particle collision rates are sufficiently high and the particle velocity distribution obeys or is close to Maxwell distribution, thus the plasma can be accurately described as a thermal fluid.
In this case, the particles in MHD are assumed to have the same macroscopic physical quantities such as the bulk velocity $\bf{u}$. 
For cases where particle collisions are insufficient, the particle velocity distribution may deviate significantly from the Maxwell distribution, where kinetic physics plays an important role. 
\citet{rinderknecht2018} pointed out that kinetic physics may contribute to the non-fluid phenomena in ICF, where differences are observed from hydrodynamic predictions of the experiments.

(ii) The MHD model of NS level is based on quasi-continuum assumption, and consequently it is only valid when Knudsen number $Kn$ (defined as the ratio of molecular mean free path $l$ to a characteristic length scale $L$) is sufficiently small. 
However, the local $Kn$ near the imposing plasma shock \cite{vidal1993ion,liu2023discrete,bond2017richtmyer} or near the perturbed interface is generally so large that it challenges the validity of the continuum assumption,\cite{mcmullen2022} resulting in inaccurate predictions of physical quantities. \cite{zhang2019entropy,zhang2019discrete,qiu2021,gan_xu2022}
Besides, the MHD model is based on the near equilibrium assumption where only the first-order Knudsen number effects are taken into account.
From this perspective, the Knudsen number $Kn$ is regarded as the ratio of relaxation time $\tau$ to the time scale $T^{Rep}$ of characteristic flow behavior.  
However, the local $Kn$ near the imposing plasma shock or near the perturbed interface is generally so large that higher order Knudsen number effects should be taken into account and consequently challenges the validity of the near-equilibrium assumption, \cite{mcmullen2022} resulting in rich and complex non-equilibrium behaviors.\cite{xu2022complex} 
More seriously, the MHD model of Euler level corresponds to the case where the local Knudsen number is zero. 
It completely neglects the viscous and heat conductive effects.

(iii) Through MHD, the non-equilibrium behaviors in the plasma system are described by the physical quantities defined in NS, such as density, temperature, pressure, viscous stress, heat conduction, etc.
These non-equilibrium behaviors can be referred to as hydrodynamic non-equilibrium (HNE).
In fact, HNE is only a small portion of the TNE,\cite{xu2022complex} where TNE is referred to as the non-equilibrium described by kinetic theory and due to that the distribution function $f$ deviates from its corresponding equilibrium distribution function $f^{eq}$.
It is foreseeable that, as the degree of discrete/non-equilibrium increases, the complexity of system behaviors increases sharply and kinetic physics becomes more and more important.
In this case, it may not be sufficient to study kinetic physics solely based on the above HNE quantities in MHD, and more TNE quantities are needed to describe the states and behaviors.\cite{xu2018discrete,xu2021progress,xu2021Progressofmesoscale,xu2021modeling}
Nowadays, TNE is attracting more attention with time, \cite{chen2017,qiu2020,qiu2021,bao2022} but is still far from being fully understood.

The second kind is some kinetic models based on non-equilibrium statistical physics, mainly based on the Boltzmann equation.
Those kinetic models can be classified into two types. The first type considers the particle collision, and the second type neglects the particle collision.
Examples of the first type are as follows: Direct simulation Monte Carlo (DSMC) \cite{bird1998recent},  
particle-in-cell (PIC) \cite{asahina2017validation}, 
hybrid fluid-PIC \citep{cai2021hybrid} and Vlasov-Fokker-Planck (VFP) method \cite{keenan2017deciphering,larroche2016ion},
 etc. Some of these methods have provided a promising solution for studying the RMI. \cite{meng2019modeling,kumar2020viscous,liu2020contribution,yan2021ion}
The second type neglects the particle collision and is based on the Vlasov equation.
It is worth noting that a considerable part of the currently adopted kinetic models ignore collisions, while researchers point out that the kinetic effects caused by particle collisions have the potential to impact the ICF.\cite{robey2004effects,rinderknecht2018,2020Cai-kinetic-effects,Yao2020Kinetic,cai2021hybrid,2021Shan-kinetic-effects}

\begin{figure*}
  \centerline{\includegraphics[width=16cm]{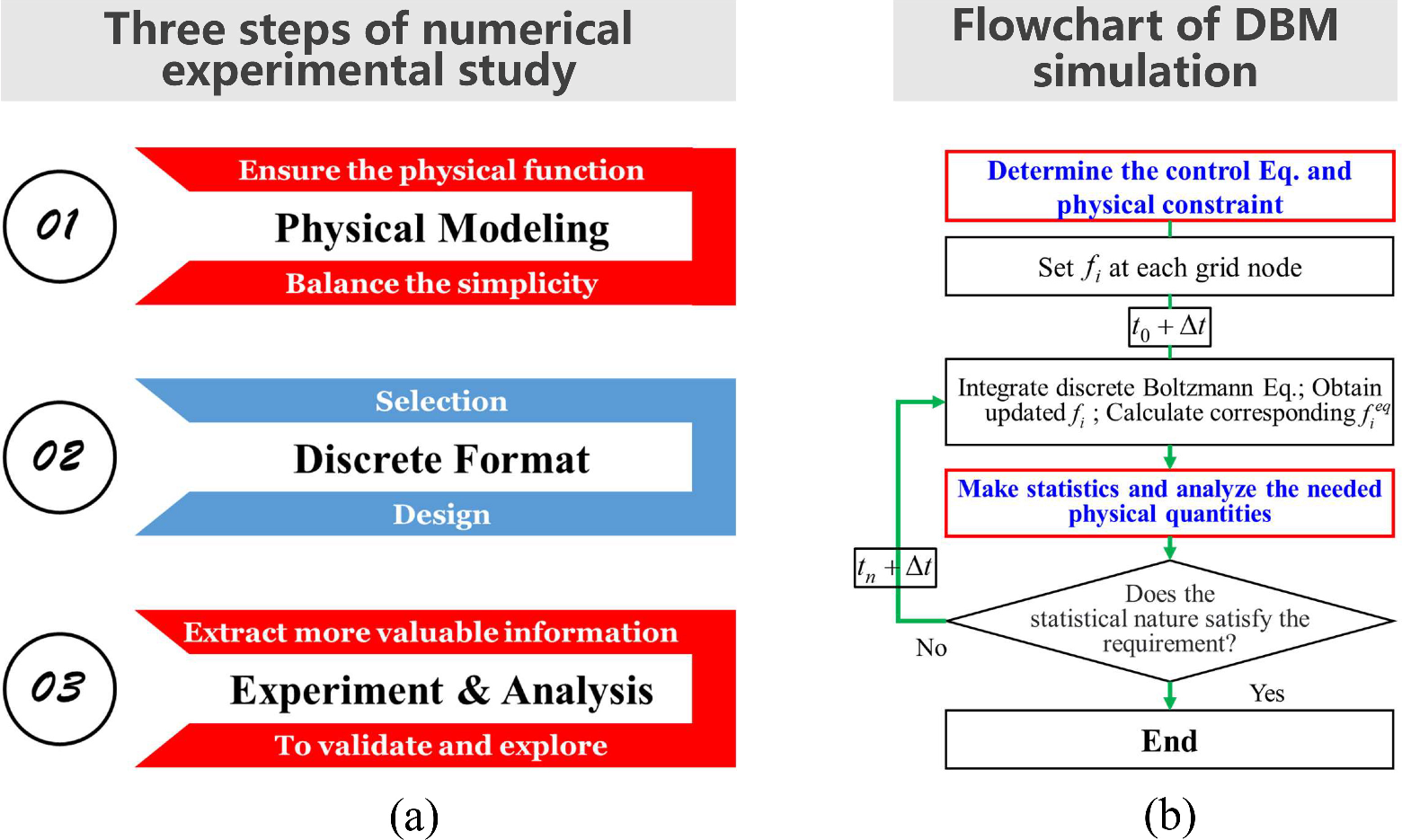}}
  \caption{Flowcharts for numerical experimental study (a) and DBM simulation study (b).}
\label{fig:1}
\end{figure*}

The content above mainly discusses how to model the physical behavior aiming to study, which prepares the evolution equation combined with necessary physical constraints for theoretical and numerical investigation.
It is well-known that numerical experiment study includes mainly three steps: (i) physical modeling, (ii) discrete format design/selection, (iii) numerical experiment and complex physical field analysis, as shown in Fig.~\ref{fig:1}(a). 
The physics group generally focuses mainly on steps (i) and (iii), while the computational mathematics group generally focuses on step (ii).  
After the numerical experiments, what we face are huge amount of data.
All the MHD and kinetic methods mentioned above are for pre-simulation services, and are not responsible for post-simulation data analysis.
Although we have made great progress in the analysis of complex physical fields, how much information do we extract from these huge amounts of data compared to the information contained in the data? Actually not optimistic. It can be said that most information is in a state of sleep.
How to extract more reliable and valuable information from these data is the key to the further development of the research.
For example, what kind of physical quantities are needed to identify and detect the discrete effects and non-equilibrium effects we mentioned above?
In order to solve the above problems, we can resort to the recently proposed discrete Boltzmann method (DBM).\cite{xu2022complex,xu2021progress,xu2021Progressofmesoscale,xu2021modeling,gan_xu2022,xu2023brief,ZHANG2023106021,zhang2023viscous}
The DBM is developed from the physical modeling branch of LBM, \cite{succi2001lattice,osborn1995lattice,swift1995lattice} thus the more accurate name of DBM is discrete Boltzmann \emph{modeling and analysis} method.

For physical modeling, the task of DBM is to build a simple physical model with sufficient physical functions to meet the problem research.
Specifically, as the degree of discrete/non-equilibrium increases, the DBM considers the insufficient particle collisions and adds more necessary non-conserved kinetic moments of distribution function $f$, which are closely related to the evolution of the system, to ensure the non-significant decrease of the system state and behavior description function.
The non-conserved moments here refer to higher-order kinetic moments other than mass, momentum, and energy conserved moments. 
The necessity of the added non-conserved kinetic moments increases with the increase of the discrete/non-equilibrium degree of the system.

For complex physical field analysis, the task of DBM is to extract more valuable information from massive data.
Specifically, the DBM uses the TNE quantities, i.e., the non-conserved kinetic moments of ($f-f^{eq}$) to describe the way and magnitude that the system deviates from its thermodynamic equilibrium state.
These TNE quantities can be selected according to practical needs, and they all describe the non-equilibrium behavior and state of the system from a certain perspective.
Since any definition of non-equilibrium strength depends on the research perspective, the degrees of TNE of these different perspectives are related and complementary, and often irreplaceable.
It is meaningful to mention that,
\emph{even for the case where the $Kn$ is small enough and consequently the NS description is reasonable, the NS shows deficiency or inconvenience for investigating some TNE behaviors, such as how the various TNE mechanisms influence the entropy production rates.}
 
Up to now, nearly all existing DBM research on hydrodynamic instability is for neutral fluids.\cite{lai2016nonequilibrium,chen2016viscosity,chen2018collaboration,
chen2020morphological,chen2022effects,li2022influence,chen2022discrete,li2022rayleigh,shan2023nonequilibrium}
In this paper, a generalized physical modeling scheme of DBM for plasma is first constructed. 
The influence of magnetic fields on HNE and TNE characteristics in the OT vortex problem and RMI are analyzed.
Some new physical insights are obtained and helpful for understanding the non-equilibrium characteristics in the plasma system.
The remainder of this paper is arranged as follows. 
The framework for constructing plasma DBM is formulated in Sec.~\ref{Discrete Boltzmann Method}.   
In Sec.~\ref{Numerical Validation}, some HNE behaviors of a number of typical benchmark problems are used to validate the first function of the DBM, and then the TNE behaviors of the Orsazg-Tang vortex problem are studied.
In Sec.~\ref{Richtmyer Meshkov instability}, the HNE and TNE behaviors of RMI with and without the initial applied magnetic field are investigated. 
At last, the conclusion and discussions are given in Sec.~\ref{Conclusion}.
\section{Discrete Boltzmann Method}
\label{Discrete Boltzmann Method}
Figure~\ref{fig:1}(b) shows a flowchart for the major steps of DBM simulation, where the two red boxes with blue words correspond to physical modeling and complex physical field analysis, respectively.
\subsection {Physical modeling} 
The origin Boltzmann equation has the ability to describe the whole flow regimes (from continuous flow to free molecular flow) and different extents of non-equilibrium effects (from quasi-equilibrium to non-equilibrium flows).
However, the high-dimensional integrals in collision term are too complicated to solve. 
To simplify, the BGK-like Boltzmann equation is adopted as 
\begin{equation}
\frac{{\partial {f}}}{{\partial t}} + {{\bf{v}}} \cdot \frac{{\partial {f}}}{{\partial {\bf{r}}}} + {\bf{a}} \cdot \frac{{\partial f}}{{\partial {\bf{v}}}} =  - \frac{1}{\tau }\left( {{f} - f^{eq}} \right),
\label{eq:1}
\end{equation}
where $f$, $f^{eq}$, $\bf{r}$, $\bf{v}$, $\bf{a}$, $t$, $\tau$ are distribution function, equilibrium distribution function, particle space coordinate, velocity, acceleration caused by external force, time and relaxation time, respectively.
The BGK model\citep{bhatnagar1954model} is adopted and the form of $f^{eq}$ is
\begin{eqnarray}
 & & {f^{eq}}\left( {\rho ,{\bf{u}},T} \right) =  \\
 &=& \rho {\left( {\frac{1}{{2\pi RT}}} \right)^{D/2}}{\left( {\frac{1}{{2\pi nRT}}} \right)^{1/2}}
\exp \left[ { - \frac{{{{\left( {{\bf{v}} - {\bf{u}}} \right)}^2}}}{{2RT}} - \frac{{{\eta ^2}}}{{2nRT}}} \right], \nonumber
\label{eq:2}
\end{eqnarray}
where $\rho$, $\bf{u}$, $T$, $R$ are density, bulk velocity, temperature, and gas constant, respectively. $D$ is the number of space dimensions, and $n$ is the number of extra degrees of freedom, with which the specific heat ratio is $\gamma=(D+n+2)/(D+n)$. $\eta$ is a free parameter that describes the energy of extra degrees of freedom including molecular rotation and vibration inside the molecules.
It should point out that the BGK-like models in various kinetic methods are actually not the original BGK-like models suitable only for quasi-static and quasi-equilibrium situations, but the revised versions modified according to the mean-field theory.\cite{xu2022complex,gan_xu2022}

In DBM, the model equation (\ref{eq:1}) needs be discretized in particle velocity space as
\begin{equation}
\frac{{\partial {f_i}}}{{\partial t}} + {{\bf{v}}_i} \cdot \frac{{\partial {f_i}}}{{\partial {\bf{r}}}} + \frac{{{\bf{a}} \cdot \left( {{\bf{u}} - {{\bf{v}}_i}} \right)}}{{RT}}f_i^{eq} =  - \frac{1}{\tau }\left( {{f_i} - f_i^{eq}} \right),
\label{eq:3}
\end{equation}
where $f_i$ and $f_i^{eq}$ are the discrete distribution function (DDF) and discrete equilibrium distribution function (DEDF), respectively. 
$i$ is the index of discrete velocities. 
Here, the distribution function $f$ in the third term on the left side of Eq.~\eqref{eq:1} is assumed to be $f^{eq}$.
\footnote{The reasons are as follows: (i) the discrete distribution function $f_i$ cannot be differentiated with respect to discrete velocity $\bf{v}$, so $f$ needs to be replaced by $f^{eq}$ before discretization, (ii) the assumption $f = f^{eq}$ suits for the cases where the non-equilibrium effects caused by force term is sufficiently weak, (iii) if the non-equilibrium effects caused by force term needs to considered, we can gradually add higher order terms according to CE multi-scale analysis $f = f^{eq} + Knf^{(1)} + Kn^2f^{(2)} + \cdots$.}
According to kinetic theory, all the system properties are described by $f$ and its kinetic moments.
For the convenience of discussion, two sets of kinetic moments are introduced as
\begin{equation}
{\bf{M}}_{m,n}\left( f \right) = \int_{ - \infty }^\infty  {{{\left( {\frac{1}{2}} \right)}^{1 - {\delta _{m,n}}}}\left( f \right)\underbrace {{{\bf{v}}}{{\bf{v}}} \cdots {{\bf{v}}}}_n{{\left( {{{\bf{v}}} \cdot {{\bf{v}}}} \right)}^{\left( {m - n} \right)/2}}d{\bf{v}}},
\label{eq:4}
\end{equation}
\begin{equation}
{\bf{M}}_{m,n}^*\left( f \right) = \int_{ - \infty }^\infty  {{{\left( {\frac{1}{2}} \right)}^{1 - {\delta _{m,n}}}}\left( f \right)\underbrace {{{\bf{v}}^*}{{\bf{v}}^*} \cdots {{\bf{v}}^*}}_n{{\left( {{{\bf{v}}^*} \cdot {{\bf{v}}^*}} \right)}^{\left( {m - n} \right)/2}}d{\bf{v}}},
\label{eq:5}
\end{equation}
where $\delta_{m,n}$ is the Kronecker delta function. 
The subscript $m,n$ represents the $m$th-order tensors contracted to $n$th-order ones. 
When $m=n$,  ${\bf{M}}_{m,n}$ and ${\bf{M}}_{m,n}^*$ are referred to as ${\bf{M}}_{m}$ and ${\bf{M}}_{m}^*$. 
${\bf{v}}^* = {{\bf{v}}} - {\bf{u}}$ represents the thermal fluctuation velocity of particles relative to bulk velocity $\bf{u}$.

For the construction of the physical model, DBM needs to ensure that the concerned system behavior cannot be changed due to discretization.
Thus, these kinetic moments must keep their value unchanged when transformed from continuous to summation form, i.e., 
\begin{equation}
\int {f{\boldsymbol{\Psi }}\left( {\bf{v}} \right)} d{\bf{v}} = \sum\limits_i {{f_i}{\boldsymbol{\Psi }}\left( {{{\bf{v}}_i}} \right)},
\label{eq:7}
\end{equation}
where ${\boldsymbol{\Psi }}\left( {\bf{v}} \right) = \left[ {1,{\bf{v}},{\bf{vv}}, \cdots } \right]$.
The left-hand side of Eq.~\eqref{eq:7} is exactly the kinetic moments used to describe the kinetic properties of the system in physical modeling.

However, there are not analytical solutions of the non-conserved kinetic moments of $f$. Eq.~\eqref{eq:7} cannot be directly used.
According to the CE multi-scale analysis, the non-conserved kinetic moments of  $f$  can be expressed as higher-order non-conserved kinetic moments of the equilibrium distribution function $f^{eq}$, and the analytical solution of $f^{eq}$ is exactly known. 
Therefore, we turn to use $f^{eq}$ to determine the most necessary physical constraints on the selection of discrete velocities.
That is,
\begin{equation}
\int {{f^{eq}}{\boldsymbol{\Psi }}'\left( {\bf{v}} \right)} d{\bf{v}} = \sum\limits_i{f_i^{eq}{\boldsymbol{\Psi }}'\left( {{{\bf{v}}_i}} \right)},
\label{eq:8}
\end{equation}
where ${\boldsymbol{\Psi }}'\left( {\bf{v}} \right) = \left[ {1,{\bf{v}},{\bf{vv}}, \cdots } \right]$ represents the higher order kinetic moments to be preserved. 

\emph{The selection of discrete velocities is the technical key of DBM, which determines the modeling accuracy.}
Eqs~(\ref{eq:7}) and (\ref{eq:8}) give the most necessary physical constraints, while the number of kinetic moments preserved could be easily determined through CE multi-scale analysis.
For example, in the DBM, five kinetic moments of $f^{eq}$ (${{\bf{M}}_0}$, ${{\bf{M}}_1}$, ${{\bf{M}}_{2,0}}$, ${{\bf{M}}_2}$, ${{\bf{M}}_{3,1}}$) are required to consider the zero-order TNE effects, seven kinetic moments of $f^{eq}$ (${{\bf{M}}_0}$, ${{\bf{M}}_1}$, ${{\bf{M}}_{2,0}}$, ${{\bf{M}}_2}$, ${{\bf{M}}_{3,1}}$, ${{\bf{M}}_3}$, ${{\bf{M}}_{4,2}})$ are required to consider the first-order TNE effects, and nine kinetic moments of $f^{eq}$ (${{\bf{M}}_0}$, ${{\bf{M}}_1}$, ${{\bf{M}}_{2,0}}$, ${{\bf{M}}_2}$, ${{\bf{M}}_{3,1}}$, ${{\bf{M}}_3}$ are required to consider the second-order TNE effects.
The details about the above kinetic moments are listed in Appendix~\ref{appA}.
Those kinetic moments could be written in matrix form as
\begin{equation}
{\bf{C}} \cdot {\bf{{{f}}^{eq}}} = {\bf{{{\hat f}}^{eq}}},
\label{eq:9}
\end{equation}
where ${\bf{C}}$, $\bf{{{{f}}^{eq}}}$, ${\bf{{{\hat f}}^{eq}}}$ are the discrete velocity polynomial, discrete equilibrium distribution function and macroscopic quantities in matrix form, respectively. 
In order to determine the specific value of ${{\bf{f}}^{eq}}$, the discrete velocity model (DVM) needs to be constructed, and the detailed discussion of the DVMs used in this paper is in Sec.~\ref{Numerical discrete schemes}.

The evolution of the electromagnetic field is described by Maxwell's equations as
\begin{equation}
\nabla  \cdot {\bf{E}} = \frac{\rho }{{{\varepsilon _0}}},
\label{eq:19}
\end{equation}
\begin{equation}
\nabla  \times {\bf{E}} =  - \frac{{\partial {\bf{B}}}}{{\partial t}},
\label{eq:20}
\end{equation}
\begin{equation}
\nabla  \cdot {\bf{B}} = 0,
\label{eq:21}
\end{equation}
\begin{equation}
\nabla  \times {\bf{B}} = {\mu _0}{\bf{j}} + \frac{1}{{{c^2}}}\frac{{\partial {\bf{B}}}}{{\partial t}}.
\label{eq:22}
\end{equation}

Several assumptions are introduced to simplify the Maxwell's equations:
(i) The Debye length is sufficiently small compared to the characteristic scale of the system, so the charge separation is ignored and the Poisson equation, that is, Eq.~(\ref{eq:19}) degenerate,
(ii) the displacement current is sufficiently small compared to the conduction current, so the second term on the right-hand side of Eq.~(\ref{eq:22}) is ignored,
(iii) the fluid is perfectly conductive and the electric conductivity is infinite, so the generalized Ohm's law is simplified as ${\bf{E}} =  - {\bf{u}} \times {\bf{B}}$.\footnote{Physically, the effects of viscous stress, heat conduction and electrical conductivity are all caused by particle collisions. The model in this paper considers a simplified case where the electrical conductivity is assumed to be infinity, which is consistent with the model of \citet{liu2018physical}. The finite electrical conductivity case will be considered in the following study.}

Physically, the divergence-free constraint Eq.~(\ref{eq:21}) is always maintained if it is initially satisfied. However, the errors caused by long-time numerical calculations can break this limit. To preserve the divergence-free constraint, the magnetic field $\bf{B}$ is represented by the magnetic potential $\bf{A}$ as ${\bf{B}} = \nabla  \times {\bf{A}}$. For two-dimensional simulation, $\bf{A}$ contains only one component $A_z$, and the evolution of ${\bf{A}}$ can be represented by the following equations,
\begin{equation}
\frac{{\partial {A_z}}}{{\partial t}} =  - {\bf{u}} \cdot \nabla {A_z} =  - {u_x}\frac{{\partial {A_z}}}{{\partial x}} - {u_y}\frac{{\partial {A_z}}}{{\partial y}}.
\label{eq:23}
\end{equation}

By solving Eq.~(\ref{eq:23}), the divergence-free constraint will automatically hold during the numerical calculation. 
In order to combine Eq.~\eqref{eq:3} and Eq.~\eqref{eq:23}, the Lorentz force is introduced in the external force term, and the acceleration $\bf{a}$ is rewritten as ${\bf{a}} = {\bf{j}} \times {\bf{B}}/\rho$, where ${\bf{j}} \times {\bf{B}}$ represent the Lorentz force as,
\begin{equation}
{\bf{j}} \times {\bf{B}} = \frac{{{\bf{B}} \cdot \nabla {\bf{B}}}}{{{\mu _0}}} - \nabla \left( {\frac{{{B^2}}}{{2{\mu _0}}}} \right),
\label{eq:24}
\end{equation}
where the first term is called the magnetic tension, while the second term is called the magnetic pressure, with which the total pressure is expressed as ${p^{*}} = p + {B^2}/\left( {2{\mu _0}} \right)$. 

Through CE multi-scale analysis, see Appendix.~\ref{appB}, the following magnetohydrodynamic model can be deduced from DBM as
\begin{equation}
\frac{{\partial \rho }}{{\partial t}} + \nabla  \cdot \left( {\rho {\bf{u}}} \right) = 0,
\label{eq:25}
\end{equation}
\begin{equation}
\frac{{\partial \left(\rho \bf{u}\right) }}{{\partial t}}+ \nabla  \cdot \left( {\rho {\bf{uu}} + p{\bf{I}}} \right) =  - \nabla  \cdot {\bf{P}}' + {\bf{j}} \times {\bf{B}},
\label{eq:26}
\end{equation}
\begin{equation}
\frac{{\partial {E_T} }}{{\partial t}} + \nabla  \cdot \left[ {\left( {{E_T} + p} \right){\bf{u}}} \right] = \nabla  \cdot \left( {\kappa \nabla T + {\bf{P}}' \cdot {\bf{u}}} \right) + \left( {{\bf{j}} \times {\bf{B}}} \right) \cdot {\bf{u}}
\label{eq:27}
\end{equation}
where ${\bf{P}}'$ is the viscous stress and $\kappa$ is the heat conductivity. 

It should be noted that the ability of recovering hydrodynamic equations is only part of the physical function of DBM.
The physical model which equivalent to DBM in physical functionality is described by the Extended Hydrodynamic Equations (EHE).
The EHE includes not only the density, momentum, and energy equations but also some of the most relevant equations of non-conserved kinetic moments.
The relationship between DBM and EHE is discussed in detail through CE multi-scale analysis in Appendix.~\ref{appC}.
Besides, the modeling method of deriving EHE from the kinetic equations is called the Kinetic Macro Modeling (KMM) method, while DBM is a Kinetic Direct Modeling (KDM) method.
This means that DBM does not need to know the specific form of corresponding EHE in simulation.
The relationship between KMM and KDM is also discussed in Appendix.~\ref{appC}.

In summary, the equations used for simulation are Eqs.~(\ref{eq:3}), Eqs.~(\ref{eq:8}), ~(\ref{eq:23}) and~(\ref{eq:24}).
\emph{As a model construction and TNE analysis method, DBM presents the basic constraints on the discrete formats.} The DBM itself does not give specific discrete formats. The specific discrete formats should be chosen according to the specific problem.
\subsection{Complex physical field analysis}
As discussed in Sec.~\ref{Introduction}, in addition to typical non-equilibrium tensity quantities such as $Kn$, $\nabla \rho$, $\nabla T$, $\nabla p$, $\cdots$, DBM uses non-conserved kinetic moments of $(f-f^{eq})$ to extract and evaluate the non-equilibrium effects.
Two kinds of TNE quantities can be defined as
\begin{equation}
{\boldsymbol{\Delta }}_{m,n} = {\bf{M}}_{m,n}\left( {f - {f^{eq}}} \right),
\label{eq:10}
\end{equation}
\begin{equation}
{\boldsymbol{\Delta }}_{m,n}^* = {\bf{M}}_{m,n}^*\left( {f - {f^{eq}}} \right),
\label{eq:11}
\end{equation}
where ${\boldsymbol{\Delta }}_{m,n}$ is referred to as central moment, and the non-equilibrium information it describes is referred to as thermo-hydrodynamic non-equilibrium (THNE).
${\boldsymbol{\Delta }}_{m,n}^*$ is referred to as non-central moment, and the non-equilibrium information it describes is referred to as TNE.
The former includes the contribution of bulk velocity $\bf{u}$, while the latter is only related to thermal fluctuation effects.

In this paper, four kinds of TNE quantities are mainly concerned (${\boldsymbol{\Delta }}_{2}^*$, ${\boldsymbol{\Delta }}_{3,1}^*$, ${\boldsymbol{\Delta }}_{3}^*$, ${\boldsymbol{\Delta }}_{4,2}^*$).
${\boldsymbol{\Delta }}_{2}^*$ is called non-organized momentum flux (NOMF), which can be regarded as generalized viscous stress.
${\boldsymbol{\Delta }}_{3,1}^*$ is called non-organized energy flux (NOEF), which can be regarded as generalized heat flux.\cite{gan_xu2022}
${\boldsymbol{\Delta }}_{3}^*$ and ${\boldsymbol{\Delta }}_{4,2}^*$ are the non-organized flux of ${\boldsymbol{\Delta }}_{2}^*$ and ${\boldsymbol{\Delta }}_{3,1}^*$, respectively. These TNE quantities contain different numbers of components. To roughly assess the strength of these TNE quantities, the average TNE strength is defined as 
\begin{equation}
D_{m,n} = \left| {{\boldsymbol{\Delta }}_{m,n}^*} \right|,
\label{eq:12}
\end{equation}
where the square of $\left| {{\boldsymbol{\Delta }}_{m,n}^*} \right|$ is equal to the sum of the squares of the individual components of $ {{\boldsymbol{\Delta }}_{m,n}^*} $. 
Similarly, the total TNE strength is defined as\cite{chen2016viscosity}
\begin{equation}
D_{T} = \sqrt {{{\left| {\boldsymbol{\Delta } _2^*} \right|}^2} + {{\left| {\boldsymbol{\Delta } _{3,1}^*} \right|}^2} + {{\left| {\boldsymbol{\Delta } _3^*} \right|}^2} + {{\left| {\boldsymbol{\Delta } _{4,2}^*} \right|}^2}}.
\label{eq:13}
\end{equation}

Considering all grids, the global average TNE strength ${{\bar D}_{m,n}}$ and ${{\bar D}_{T}}$ are defined as
\begin{equation}
{{\bar D}_{m,n}} = \frac{1}{{{N_x} \times {N_y}}}\sum {{D_{m,n}}},
\label{eq:14}
\end{equation}
\begin{equation}
{{\bar D}_{T}} = \frac{1}{{{N_x} \times {N_y}}}\sum {{D_{T}}},
\label{eq:15}
\end{equation}
where $N_x$ and $N_y$ are the grid numbers in $x$ and $y$ direction, respectively.

\emph{Entropy generation rate is an important parameter concerned in many fields related to compression science, such as shock wave physics, ICF, and aerospace.} 
Through TNE quantities, the total entropy production rate is defined as \citep{zhang2016kinetic,zhang2019entropy},
\begin{equation}
{{\dot S}_{b}} = \frac{{d{S_b}}}{{dt}} = \int {\left( {{\boldsymbol{\Delta }}_{3,1}^* \cdot \nabla \frac{1}{T} - \frac{1}{T}{\boldsymbol{\Delta }}_2^*:\nabla {\bf{u}}} \right)d{\bf{r}}}.
\label{eq:16}
\end{equation}
where ${{\dot S}_{b}}$ can be divided by the part denoted by NOMF and NOEF as,
\begin{equation}
{{\dot S}_{NOMF}} = \int { - \frac{1}{T}{{\Delta }}_2^*:\nabla {\bf{u}}d{\bf{r}}},
\label{eq:17}
\end{equation}
\begin{equation}
{{\dot S}_{NOEF}} = \int {{\boldsymbol{\Delta }}_{3,1}^* \cdot \nabla \frac{1}{T}d{\bf{r}}}.
\label{eq:18}
\end{equation}

To give a complete description of a complex physical field, the non-equilibrium intensity vector $\bf{D}$ can be further introduced.
Each component of $\bf{D}$ is a non-equilibrium intensity parameter (including various TNE quantities and other necessary HNE quantities). \cite{xu2021modeling,zhang2022discrete}
For example, in this paper, a non-equilibrium intensity vector is introduced as
\begin{eqnarray}
 \mathbf{D} = \left\{  {D}_{T}, {D}_{2},{D}_{3,1},{D}_{3},{D}_{4,2}, {{\bar D}_{T}},{{\bar D}_{2}},{{\bar D}_{3}},{{\bar D}_{3,1}}, {{\bar D}_{4,2}}, \right. \\ \nonumber
 \left. \left|{\dot S_{NOMF}} \right|, \left|{\dot S_{NOEF}} \right|, \left| {\nabla \rho } \right|, \left| \nabla T \right|, \left| \nabla p \right|, Kn, \cdots \right\}
\label{eq:D}
\end{eqnarray}

Since there exist infinite perspectives on non-equilibrium behaviors, the more perspectives are chosen, the more accurate the description of the system.
The non-equilibrium intensity vector can be used to open a phase space, which gives an intuitive geometric correspondence to the non-equilibrium state of the system.
\section{Numerical Validation and Investigation}
\label{Numerical Validation}
In this section, the numerical discrete schemes used in DBM simulation are introduced, including the spatial, temporal, and particle velocity space discrete schemes.
Then, the non-dimensionalization method is discussed.
Finally, a number of typical benchmark problems such as sod shock tube, thermal Couette flow, and the compressible Orszag-Tang (OT) vortex problem are simulated, where the hydrodynamic behaviors are used to validate the plasma DBM. 
Here, we only show the results of the OT vortex problem. 
\subsection{Numerical discrete schemes}
\label{Numerical discrete schemes}
\begin{figure}
  \centerline{\includegraphics[width=9cm]{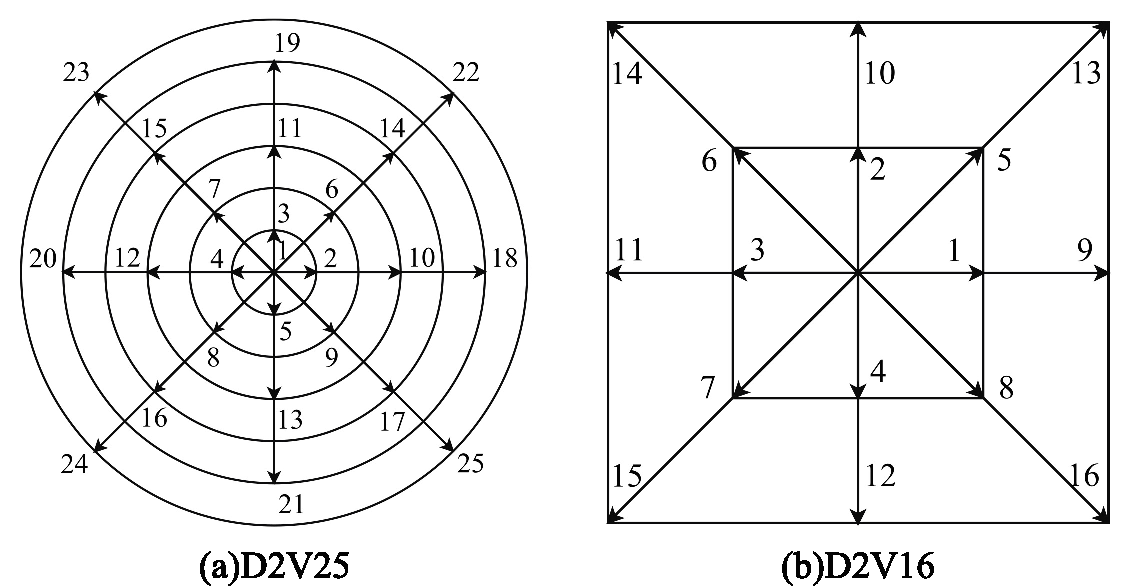}}
  \caption{Schematics of two kinds of discrete velocity models used in the present paper. (a)D2V25, (b)D2V16. The numbers in the figures represent the index of discrete velocities.}
\label{fig:1a}
\end{figure}
\begin{figure*}
\centerline{\includegraphics[width=17cm]{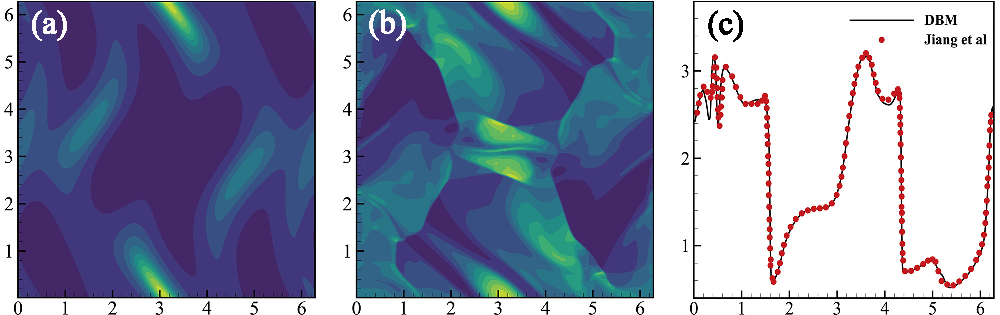}}
\caption{DBM results of Orszag-Tang vortex problem. (a) Pressure contour at time $t=0.5$ of DBM results. (b) Pressure contour at time $t=3$ of DBM results. (c) Pressure distributions along $y=0.625\pi$ at $t=3$. The solid line represent the DBM result while the red dots represent the result of \citet{jiang1999}.}
\label{fig:2}
\end{figure*}

 
In this paper, the first-order forward Euler finite difference scheme and the second-order non-oscillatory nonfree dissipative (NND) scheme are used to discrete spatial and temporary derivatives in Eq.~(\ref{eq:3}) and Eq.~(\ref{eq:23}), respectively. The second-order central difference scheme is used to discretize the space derivative in Eq.~(\ref{eq:24}).
Two two-dimensional DVMs are used in this paper.  
The DVM used to simulate the benchmark problems has $25$ discrete velocities, as shown in Fig.~\ref{fig:2}(a).
The DVM used to simulate the RMI has $16$ discrete velocities, as shown in Fig.~\ref{fig:2}(b).

The mathematical expressions of these two kinds of DVM are as follows,
\begin{eqnarray}
\renewcommand{\arraystretch}{1.5}
&&\text{D2V25}:\\ \nonumber
&&({v_{ix}},{v_{iy}}) = \left\{ \begin{array}{l}
0,\qquad \qquad \qquad \qquad \qquad \qquad  i = 1,\\
c[\cos \frac{{(i - 2)\pi }}{2},\sin \frac{{(i - 2)\pi }}{2}],{\rm{ \qquad    }}i = 2 - 5,\\
2c[\cos \frac{{(2i - 3)\pi }}{4},\sin \frac{{(2i - 3)\pi }}{4}],{\rm{ \quad     }}i = 6 - 9,\\
3c[\cos \frac{{(i - 10)\pi }}{2},\sin \frac{{(i - 10)\pi }}{2}],{\rm{      }}i = 10 - 13,\\
4c[\cos \frac{{(2i - 11)\pi }}{4},\sin \frac{{(2i - 11)\pi }}{4}],{\rm{    }}i = 14 - 17,\\
5c[\cos \frac{{(i - 18)\pi }}{2},\sin \frac{{(i - 18)\pi }}{2}],{\rm{   }}i = 18 - 21,\\
6c[\cos \frac{{(2i - 19)\pi }}{4},\sin \frac{{(2i - 19)\pi }}{4}],{\rm{  }}i = 22 - 25,
\end{array} \right.
\label{eq:28}
\end{eqnarray}
\begin{eqnarray}
\renewcommand{\arraystretch}{1.5}
\text{D2V16}:\left( {{v_{ix}},{v_{iy}}} \right) = \left\{ \begin{array}{l}
cyc:c\left( { \pm 1,0} \right){\rm{,  \quad   \quad  \;    1}} \le i \le 4,\\
c\left( { \pm 1, \pm 1} \right){\rm{,   \quad   \quad  \quad  \quad     1}} \le i \le 8,\\
cyc:2c\left( { \pm 1,0} \right){\rm{,   \quad \quad  9}} \le i \le 12,\\
2c\left( { \pm 1, \pm 1} \right){\rm{,  \quad \quad \quad \;  \;    13}} \le i \le 16,
\end{array} \right.
\label{eq:29}
\end{eqnarray}
where `cyc' represents the indicates cyclic permutation. $c$ is a free parameter to optimize the properties of the DVM.

It's worth noting that the standard Lattice Boltzmann Method (LBM) utilizes the "propagation + collision" concept with the discrete velocity dictating the direction of particle movement. \cite{succi2001lattice}
In DBM, although the discrete velocity remains in use, the "propagation + collision" image is no longer present.
The function of DVM is to ensure that the physical constraints described by Eqs.~\eqref{eq:7} and \eqref{eq:8} strictly hold. \cite{xu2022complex}
Therefore, the construction of DVM in DBM is very flexible, which is based on comprehensive consideration of physical symmetry, computational efficiency, etc.
\subsection{Non-dimensionalization}
In this work, all physical quantities are non-dimensionalized using reference density $n_0$, temperature $T_0$, and length $L_0$ as
\begin{equation}
\hat \rho  = \frac{\rho }{{{\rho _0 }}}, \quad  \hat T = \frac{T}{{{T_0 }}}, \quad \left( {\hat t,\hat \tau } \right) = \frac{{\left( {t,\tau } \right)}}{{L_0/ \sqrt {R{T_0 }} }}, \quad {\hat r_\alpha } = \frac{{{r_\alpha }}}{{{L_0 }}},  \nonumber
\label{eq:30}
\end{equation} 
\begin{equation}
\hat P = \frac{P}{{{\rho _0}{R }T_0}}, \quad \left( {\hat f,{{\hat f}^{eq}},{{\hat f}_i},\hat f_i^{eq}} \right) = \frac{{\left( {f,{f^{eq}},{f_i},f_i^{eq}} \right)}}{{{\rho _0 }{{\left( {R{L_0 }} \right)}^{ - 1}}}},  \nonumber
\label{eq:31}
\end{equation}
\begin{equation}
\left( {{{\hat v}_\alpha },{{\hat c}_\alpha },{{\hat u}_\alpha }} \right) = \frac{{\left( {{v_\alpha },{c_\alpha },{u_\alpha }} \right)}}{{ { \sqrt {R{T_0 }}  } }}, \quad \hat B = \frac{{B{L_0 }}}{{{t_0 }}}\sqrt {{\mu _0}{\rho _0 }}, \nonumber
\label{eq:32}
\end{equation}
\begin{equation}
\hat \kappa  = \frac{\kappa }{{{\rho _0 }R{L_0 }\sqrt {R{T_0 }} }}, \quad \hat \mu  = \frac{\mu }{{{\rho _0 }{L_0 }\sqrt {R{T_0 }} }}, \nonumber
\label{eq:33}
\end{equation}
where $\kappa$ and $\mu$ are the heat conduction and viscosity coefficient, respectively.
The variables on the left-hand side with the symbol ``$\land$'' are dimensionless, while the variables on the right-hand side without the symbol ``$\land$'' are dimensional.
For the sake of simplicity, the ``$\land$'' of dimensionless variables is omitted in the following sections.

The $Kn$ in BGK model is defined as,\cite{xu2022complex}
\begin{equation}
Kn = \frac{{{\lambda _{BGK}}}}{L} = \frac{\tau }{L}\sqrt {\frac{{8RT}}{\pi }}  = \frac{{\hat \tau }}{{\hat L}}\sqrt {\frac{{8\hat T}}{\pi }},
\label{eq:34B}
\end{equation}
where $\hat L$ is the non-dimensional characteristic length scale.
\subsection{Orszag-Vortex problem}
This problem was first introduced by Orszag and Tang in 1979.\citep{orszag1979small} Due to the complex vortex and shock wave structures generated during evolution, this problem has been widely used to demonstrate the validity of new models. 
Here, the initial configuration and conditions are identical to \citet{jiang1999}, as follows,
\begin{eqnarray}
\begin{array}{l}
\rho \left( {x,y,0} \right) = {\gamma ^2},  \qquad \qquad \qquad{\rm{          }}{v_x}\left( {x,y,0} \right) =  - \sin y,{\rm{   }}\\
{v_y}\left( {x,y,0} \right) = \sin x,            \qquad \qquad \quad{\rm{       }}p\left( {x,y,0} \right) = \gamma ,\\
{B_x}\left( {x,y,0} \right) =  - \sin y,        \qquad \qquad{\rm{   }}{B_y}\left( {x,y,0} \right) = \sin 2x,
\end{array}
\label{eq:34}
\end{eqnarray}
where $\gamma $ is equal to $5/3$. 
The simulation is performed on a computational domain of size $\left[ {0,2\pi } \right] \times \left[ {0,2\pi } \right]$, which has been divided into $N_{x} \times N_{y}= 400 \times 400$ mesh-cells. The D2V25 model (shown in Fig.~\ref{fig:1}) is used to discretize the particle velocity space, where $c = 0.6$, ${\eta _0} = 0.4$ for $i=2,\cdots,5$, ${\eta _0} = 0.8$ for $i=6,\cdots,9$ and ${\eta _0} = 0$ for others. 
The time step is $\Delta t = 5 \times {10^{ - 4}}$, the space step is $\Delta x = \Delta y = 5 \times {10^{ - 3}}\pi $, the relaxation time is $\tau  = 1 \times {10^{ - 3}}$, and the number of extra degree of freedom is $n = 1$. Moreover, the periodic boundaries are applied in both the $x$ and $y$ direction.

Figure~\ref{fig:2} shows the numerical results obtained by DBM. 
From Fig.~\ref{fig:2}(b), it is observed that the system is very complex, forming a variety of vortex and shock wave structures, especially in the center of the flow field. 
By comparing the DBM results in Figs.~\ref{fig:2}(a)-(b)  with the MHD results in Figs. (12)-(14)  given by \citet{jiang1999}
, it was found that the two sets of results are in good agreement. 
For quantitative analysis, the pressure distribution along $y=0.625\pi$ 
at time $t=3$ was extracted and plotted together with the result of \citet{jiang1999} in Fig.~\ref{fig:2}(c). 
Although the DBM results from $x=0.2$ to $x=0.4$ were slightly lower, the remaining results matched those of \citet{jiang1999}, thus validating the new DBM.
\begin{figure*}
  \centerline{\includegraphics[width=17cm]{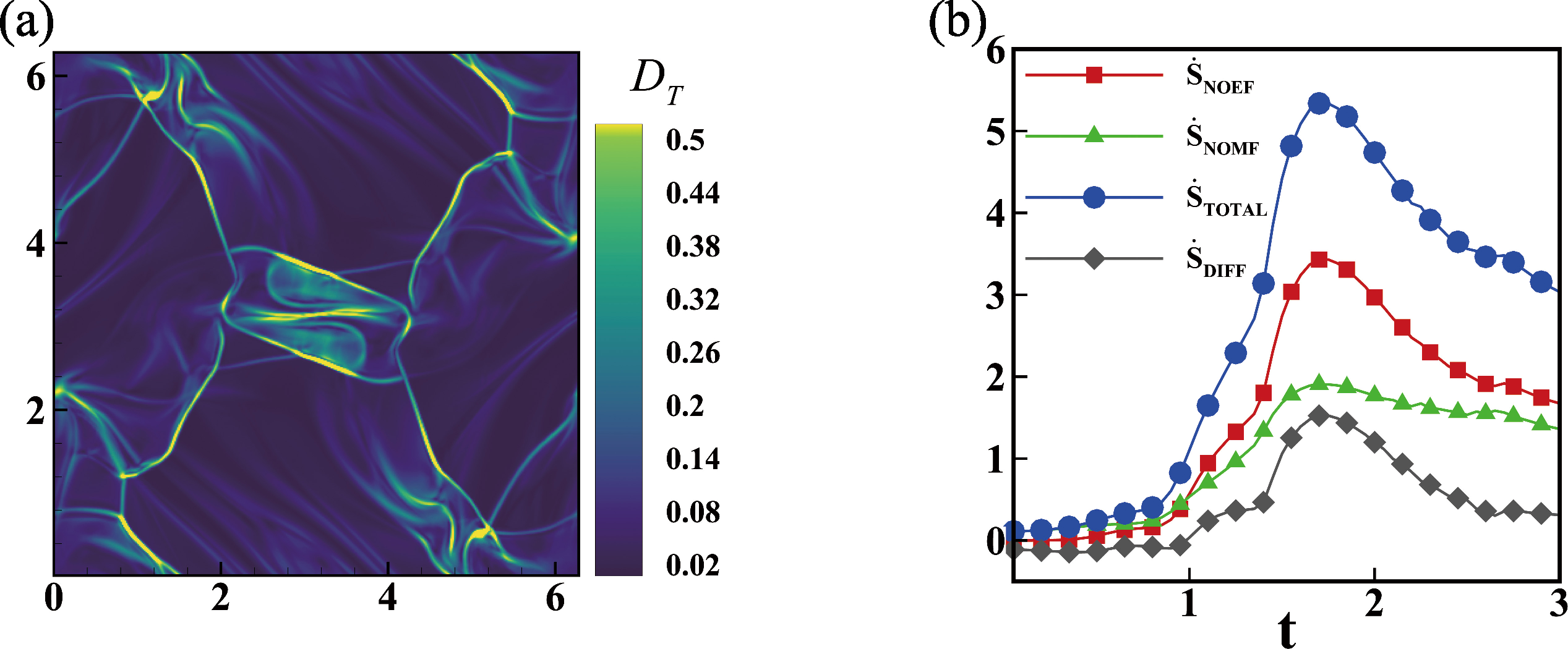}}
  \caption{TNE effects of Orszag-Tang vortex problem. (a) Contour of total TNE strength $D_{T}$ at $t=3$, (b) evolution of three kinds of entropy production rate with time. The red line with square symbol corresponding to ${{\dot S}_{NOEF}}$. The green line with delta symbol corresponding to ${{\dot S}_{NOMF}}$. The blue line with circle symbol corresponding to the total entropy production rate. The gray line with diamond symbol corresponding to the difference between ${{\dot S}_{NOEF}}$ and ${{\dot S}_{NOMF}}$.}
\label{fig:3}
\end{figure*}

It is worth noting that the kinetic behaviors of OT turbulence evolution are still not well-understood.
Additionally, the TNE characteristics were further extracted and plotted in Fig.~\ref{fig:3}.
Figure~\ref{fig:3}(a) shows the contour map of total TNE strength ${D_{T}}$  at $t=3$. 
It is found that the TNE effects are most pronounced in the shock front where high physical gradients exist. 
For the rest of the flow field, the TNE strength is weak, indicating that the system is close to the local thermodynamic equilibrium state.
Figure~\ref{fig:3}(b) shows the evolution of four kinds of entropy production rates with time. 
It can be seen that both the four kinds of entropy production rates show two stages effect: before $t=1.7$, the entropy production rates keep increasing with time; After $t=1.7$, the entropy production rates reach the peak value and then keep decreasing with time. 
In fact, there is a positive correlation between the entropy production rate and the physical quantity gradient. 
Before $t=1.7$, the evolution of the flow field is dominated and the velocity and temperature gradients keep increasing, which leads to the increase of entropy production rates. After $t=1.7$, the dissipative effects are dominated, resulting in the decrease of gradients and entropy production rates. Besides, in the early stage, it is found that the ${{\dot S}_{NOEF}}$ increases from $0$, while the ${{\dot S}_{NOMF}}$ increases from about $0.1$. This is because the initial temperature field is uniform and there exists no temperature gradient. For the initial velocity field, the initial velocity gradient induces a momentum exchange and leads to an initial entropy production rate ${{\dot S}_{NOMF}}$. Besides, as the flow field develops, the entropy production rate ${{\dot S}_{NOEF}}$ exceeds ${{\dot S}_{NOMF}}$ after $t=1$ and is always larger than ${{\dot S}_{NOMF}}$ in the subsequent evolution, indicating that the entropy production caused by NOEF is in dominant in the late stage of OT vortex problem.

\emph{From the perspective of compression science, generally, the entropy production rate represents the compression difficulty.} In the process of OT evolution, the difficulty of compression is divided into stages: in the first stage, the entropy production rate and compression difficulty increase with time; while in the second stage, the entropy production rate and compression difficulty decrease with time.
\section{Richtmyer-Meshkov instability}
\label{Richtmyer Meshkov instability}
In this section, the RMI induced by a shock wave passing through a heavy/light density interface is simulated by using the current DBM. 
The effects of different initial magnetic field ${B_0}$ on the evolution of RMI and the subsequent re-shock process are carefully investigated. 
This section consists of three subsections. In the first subsection, the initial configuration of RMI is given. In the next subsection, the HNE and TNE effects of RMI without magnetic fields are investigated. In the last subsection, the HNE and TNE effects of RMI with different initial magnetic fields are carefully investigated. Meanwhile, the entropy production rates with different initial magnetic fields are calculated and analyzed. 
As a preliminary application of related research work, this work adopts a first-order model.
The reason is as follows: (i) in the case of low-intensity shocks, the first-order model is sufficient, and (ii) the results of these low-level models provide a technical and cognitive basis for the next step of higher-intensity shock scenarios.
\subsection{Flow field Settings}
\label{settings}
\begin{figure}
  \centerline{\includegraphics[width=9cm]{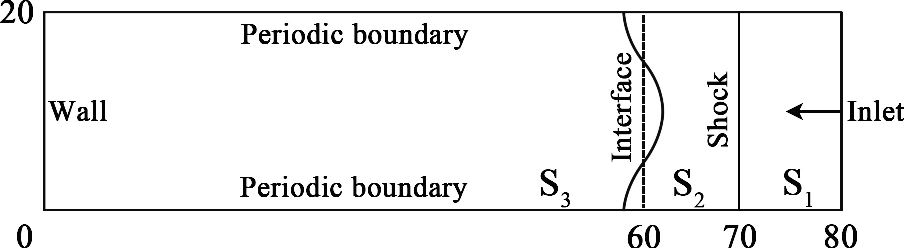}}
  \caption{Schematic of initial configuration of Richtmyer-Meshkov instability.}
\label{fig:4}
\end{figure}

Figure~\ref{fig:4} shows the initial flow field of RMI. The length and height of the two-dimension computation domain are 20 and 80, respectively. 
The computation domain is divided into $N_x \times N_y = 200 \times 800$ mesh-cells. 
Initially, a sinusoidal perturbation interface is located at $y = 3N_{y}/4$, with wavelength $\lambda  = d$, where $d=20$ is equal to the length of the computation domain. 
The amplitude is set to be ${y_0} = 0.1d$, which corresponds to a small perturbation case. 
The initial shock wave is located at $y = 7N_{y}/8$, and the physical quantities on the left and right sides of the shock wave were connected by the Rankine–Hugoniot conditions.
Thus, the shock wave and perturbed interface separate the domain into three regions as ${S_1}$, ${S_2}$, and ${S_3}$. ${S_1}$ is the area that has been compressed by the passed shock wave. ${S_2}$ is the region with high density and ${S_3}$ is the region with light density. Furthermore, the periodic boundary conditions are applied to the boundaries in the $x$ direction, and the free inflow and solid boundary conditions are applied to the boundaries at the top and bottom of the $y$ direction, respectively. 
The initial hydrodynamic quantities are as follows,
\begin{eqnarray}
\left\{ \begin{array}{l}
{\left( {\rho ,{u_x},{u_y},p} \right)_{S1}} = \left( {3.1304,0,-0.28005,2.1187} \right)\\
{\left( {\rho ,{u_x},{u_y},p} \right)_{S2}} = \left( {2.3333,0,0,1.4} \right)\\
{\left( {\rho ,{u_x},{u_y},p} \right)_{S3}} = \left( {1,0,0,1.4} \right)
\end{array} \right.
\label{eq:35}.
\end{eqnarray}

With the above initial quantities, a shock wave with $\rm{Ma}= 1.2$  is generated to impact the sinusoidal perturbation interface. The Atwood number is $\rm{At}=0.4$. The D2V16 model in Fig.~\ref{fig:1a} is adopted to discrete the particle velocity space, and the parameters are $c=1$, ${\eta _0} = 5$ for $i=1,\cdots,4$ and ${\eta _0} = 0$ for others. The other parameters of DBM are as follows: space step $\Delta x = \Delta y = 1 \times {10^{ - 1}}$, time step $\Delta t = 1 \times {10^{ - 3}}$, relaxation time $\tau = 1 \times {10^{ - 3}}$ and the number of extra degrees of freedom $n=3$, i.e., the specific heat ratio $\gamma=1.4$. The maximum initial CFL number is
\begin{equation}
{C_0} = \frac{{\max \left( {\left| {{u_y}} \right| + {c_{\max }}} \right) \cdot \Delta t}}{{\Delta x}} = 0.0228,
\label{eq:C}
\end{equation}
where ${c_{\max }} = 2$ is the maximum velocity in DVM.
\subsection{Grid convergence test}
\begin{figure*}
  \centerline{\includegraphics[width=16cm]{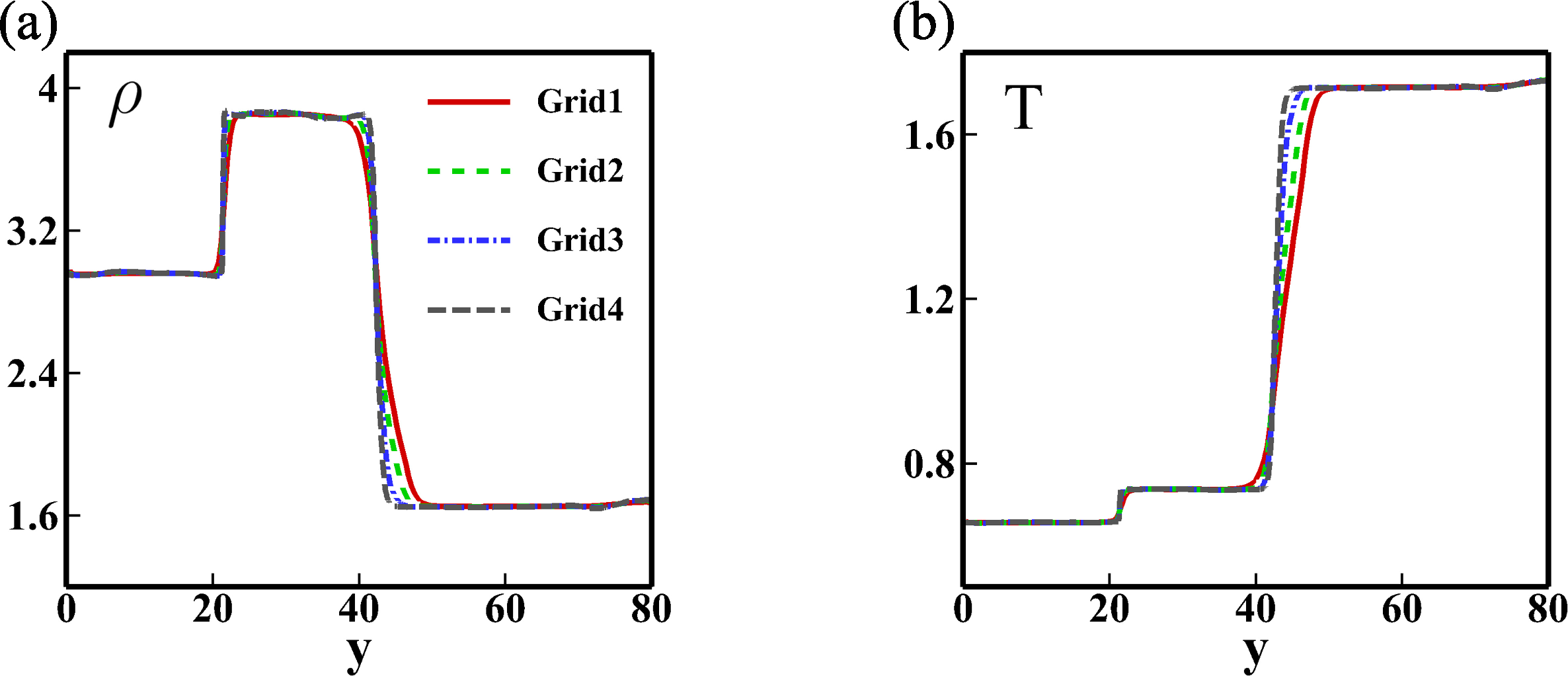}}
  \caption{Distribution of physical quantities along $x=10$ of four kinds of grid at $t=100$. (a)Density, (b)Temperature.}
\label{fig:appC}
\end{figure*}
In order to verify the effectiveness of numerical resolution, a grid convergence test is performed. Four kinds of grid numbers are selected as Grid1: $N_x \times N_y = 50 \times 200$, Grid2: $N_x \times N_y = 100 \times 400$, Grid3: $N_x \times N_y = 200 \times 800$ and Grid4: $N_x \times N_y = 400 \times 1600$. The corresponding space steps are $\Delta x = \Delta y = 4 \times {10^{ - 1}}$, $\Delta x = \Delta y = 2 \times {10^{ - 1}}$, $\Delta x = \Delta y = 1 \times {10^{ - 1}}$, $\Delta x = \Delta y = 5 \times {10^{ - 2}}$, respectively. The other calculation parameter settings of DBM are: time step $\Delta t = 1 \times {10^{ - 1}}$, relaxation time $\tau = 1 \times {10^{ - 3}}$ and the number of extra degrees of freedom $n=3$, i.e., the specific heat ratio $\gamma=1.4$. Figure~\ref{fig:appC} shows the distribution of density and temperature along $x=N_x/2$. It can be seen that the results of $200 \times 800$ and $400 \times 1600$ meshes are quite identical. By comprehensively considering numerical resolution and computational cost, the $200 \times 800$ meshed are selected for calculation and analysis in this paper.
\subsection{RMI without magnetic field}
In this section, the HNE and TNE effects of RMI without magnetic field are investigated. 
The magnetic field strength $B_0 = 0.0$, and the other parameters are consistent with those in Sec.~\ref{settings}.
\subsubsection{HNE characteristics}
\begin{figure*}
  \centerline{\includegraphics[width=17cm]{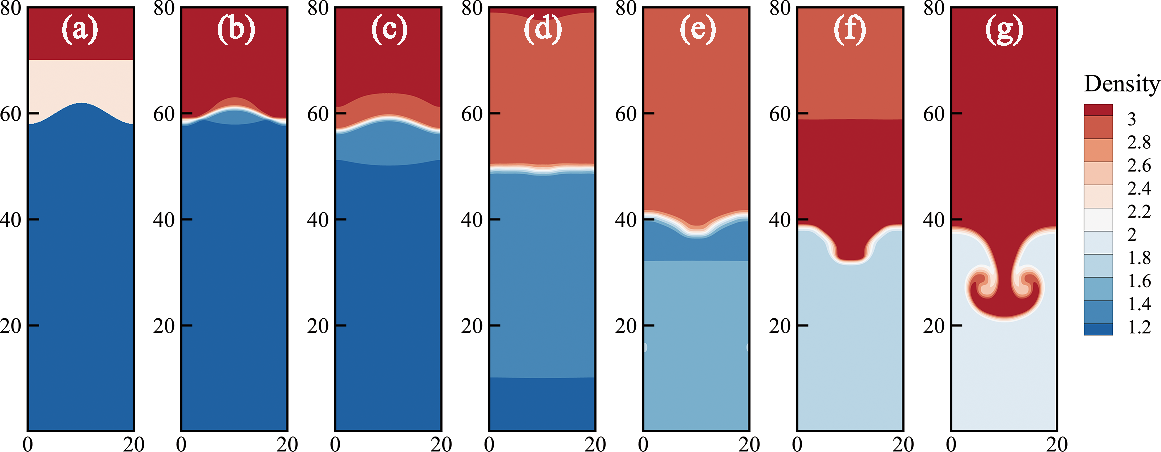}}
  \caption{Density contours in the evolution of RMI without initial applied magnetic field at different times: (a)$t=0$, (b)$t=10$, (c)$t=15$, (d)$t=40$, (e)$t=70$, (f)$t=100$, (g)$t=250$.}
\label{fig:5}
\end{figure*}
Figure~\ref{fig:5} shows the density evolution in RMI via snapshots at times $t = 0$,$10$,$15$,$40$,$70$,$100$,$250$. At $t=10$, the shock wave hits the interface and interacts with the interface. At $t=15$, the shock wave passes through the interface, forming a reflected rarefaction wave propagating upward to the top boundary and a transmission shock wave propagating downward to the bottom boundary. Meanwhile, the perturbation amplitude gradually decreases to zero with the motion of the interface. Then, the interface reverses and the perturbation amplitude grows again, accompanied by the reversal of the initial perturbation interface peak and valley. At $t=40$, the transmission shock wave moves to the bottom solid wall and reflects. At $t=70$, the reflected transmission shock wave passes through the interface again, forming a reflected rarefaction wave to light fluids and a transmission shock wave to heavy fluids. Under the action of the secondary shock (reflected transmission shock wave), the speed of interface development is greatly accelerated and the speed difference on both sides of the interface increases rapidly. Then, the Kelvin-Helmholtz instability (KHI) appears at the head of the spike, eventually forming a mushroom-like structure. During the interface evolution, the light and heavy fluids continuously mix with each other.
\subsubsection{TNE effects}
\begin{figure*}
  \centerline{\includegraphics[width=17cm]{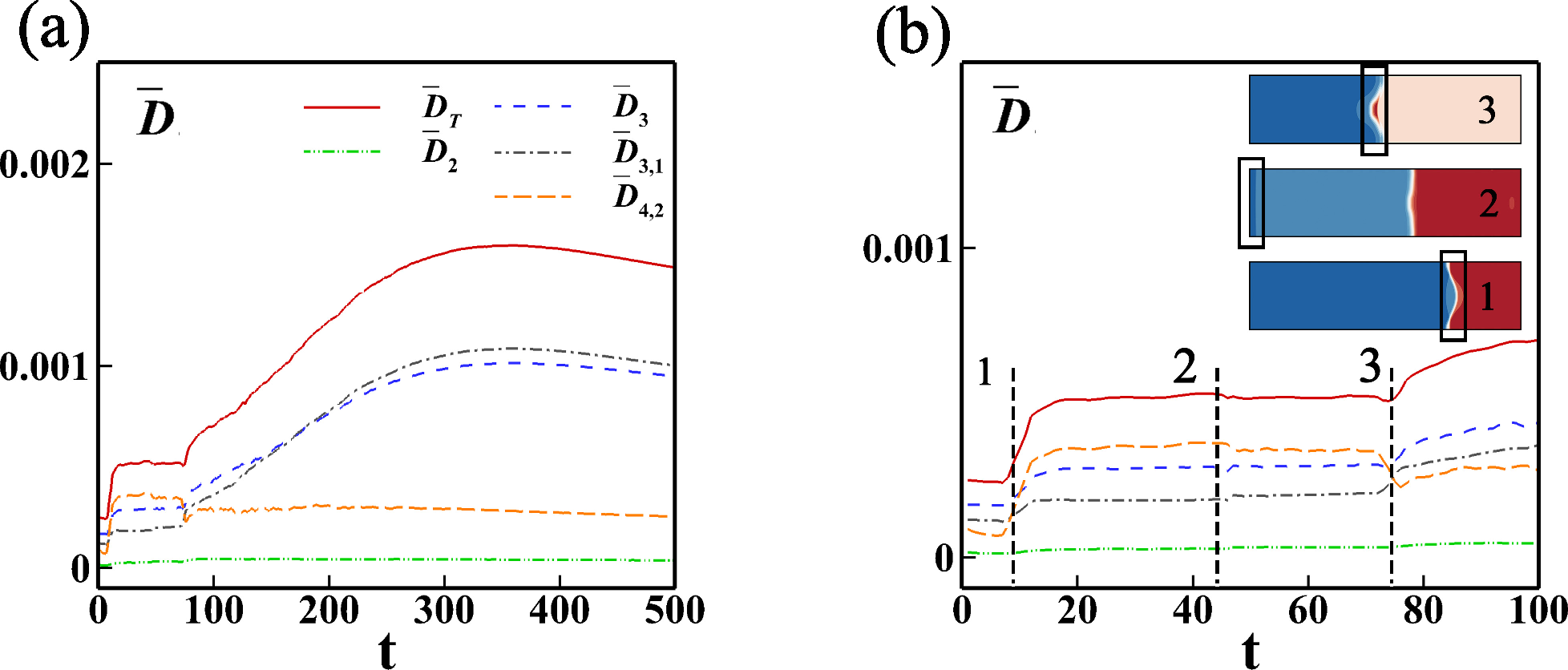}}
  \caption{Contours of global average TNE quantities ${\bar D_{T}}$, ${\bar D_{2}}$, ${\bar D_{3}}$, ${\bar D_{3,1}}$ and ${\bar D_{4,2}}. $ (a) The time ranges from $t=0$ to $t=500$, (b) the time ranges from $t=0$ to $t=100$.}
\label{fig:7}
\end{figure*}

Figure~\ref{fig:7}(a) shows the global average TNE strength ${\bar D_{T}}$, NOMF ${\bar D_2}$, NOEF ${\bar D_{3,1}}$, the flux of NOMF ${ \bar D_{3}}$ and the flux of NOEF ${\bar D_{4,2}}$ from $t=0$ to $t=500$. It can be found that the strength of ${\bar D_2}$ is much lower than other TNE effects, and the trend of ${\bar D_{T}}$ is basically the same as that of ${\bar D_{3,1}}$.
This means that the TNE effects caused by viscosity are weaker than that caused by heat conduction.
The TNE strength ${\bar D_{T}}$ of the system basically increases with time before about $t=320$, while decreases with time after $t=320$. In fact, there exist two competition mechanisms. Before $t=320$, the evolution of RMI is at the linear and weak nonlinear development stage. 
With the development of the interface, the physical quantities gradient near the interface increases rapidly, and the non-equilibrium region augments, which causes the increasing of ${\bar D_{T}}$. After  $t=320$, the evolution of RMI is at the late stage of nonlinear development. 
At this time, the dissipation effects such as viscosity and heat conduction are dominant.
This increase the degree of mixing near the interface and reduce the gradient of physical quantities, causing the decrease of ${\bar D_{T}}$. 
Besides, from the red line $({\bar D_{T}})$, three key time points are marked, which are shown in the enlarged picture Fig.~\ref{fig:7}(b).

Figure~\ref{fig:7}(b) shows the the global average TNE strength ${\bar D_{T}}$, NOMF ${\bar D_2}$, NOEF ${\bar D_{3,1}}$ and the flux of NOEF ${\bar D_{4,2}}$ from $t=0$ to $t=100$. 
It is found that, before point 1, TNE effects decrease very slowly with time due to dissipation effects. 
At point 1, when the shock wave passes through the interface, the distribution function near the interface greatly deviate from the local thermodynamic equilibrium state.
After point 1, the dissipative effects tend to reduce TNE effects, while the high $Kn$ makes it difficult to recover to the local thermodynamic equilibrium state.
Thus, all TNE effects increase.
At point 2, the transmission shock wave hits the boundary and reflects. 
In this process, the variation of the macroscopic physical quantities gradients leads to a fluctuation of TNE quantities.
At point 3, the reflected transmission shock wave hits the interface again. 
Since the direction of the shock wave is opposite to that of the first time, not all TNE quantities increase. The flux of NOEF, i.e., ${\bar D_{4,2}}$ decreases, while the rest of the TNE quantities increase. Because the interface has reversed when the shock wave hits the interface for the second time, the vorticity generated is further deposited at the interface after point 3.
This leads to the accelerated development of the interface, the expansion of the non-equilibrium area, and the continuous increase of all TNE quantities.
\begin{figure*}
  \centerline{\includegraphics[width=17cm]{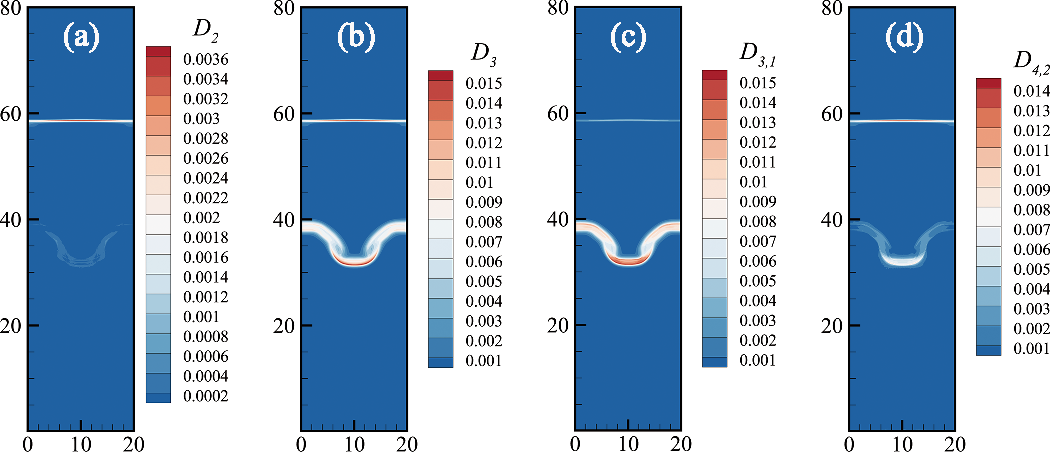}}
  \caption{Contours of different TNE quantities of RMI at $t=100$. (a)$D_2$, (b)$D_3$, (c)$D_{3,1}$, (d)$D_{4,2}$}
\label{fig:8}
\end{figure*}

Figure~\ref{fig:8} shows the contours of TNE quantities $D_2$, $D_3$, $D_{3,1}$, $D_{4,2}$ at $t=100$, at which time the reflected transmission shock wave has passed the perturbed interface. 
From Fig.~\ref{fig:8}(a), it is found that $D_2$ mainly distributes at the shock wavefront and the perturbed interface, and the intensity at the shock wavefront is much higher than that at the perturbed interface. 
For the rest of the flow field, the $D_2$ is almost zero. Therefore, $D_2$ can be used to capture the location of the shock wave during RMI. 
Due to the strong momentum transport and shear effects caused by the large density and velocity gradients, the $D_2$ is most remarkable at the shock wavefront. 
Figure~\ref{fig:8}(c) shows the distribution of $D_{3,1}$. 
It reaches the maximum value at the perturbed interface. 
Thus, $D_{3,1}$ can be used to capture the evolution of the interface and the amplitude. 
In this case, the temperature gradient near the perturbed interface is bigger than that near the shock wavefront, causing strong energy transport. 
Thus, the $D_{3,1}$ is most remarkable at the perturbed interface. Though $D_2$ and  $D_{3,1}$ can be used to identify the shock wavefront and perturbed interface respectively, they cannot be used to identify both of them. Compared with $D_2$ and  $D_{3,1}$, $D_3$ and $D_{4,2}$ are more suitable. $D_3$ provides the most distinguishable interfaces, which is also proved by \citet{zhang2019discrete}. During the evolution of RMI, both of these four TNE quantities can be used as the supplement of the macroscopic quantities contours of the flow field.
\begin{figure*}
  \centerline{\includegraphics[width=17cm]{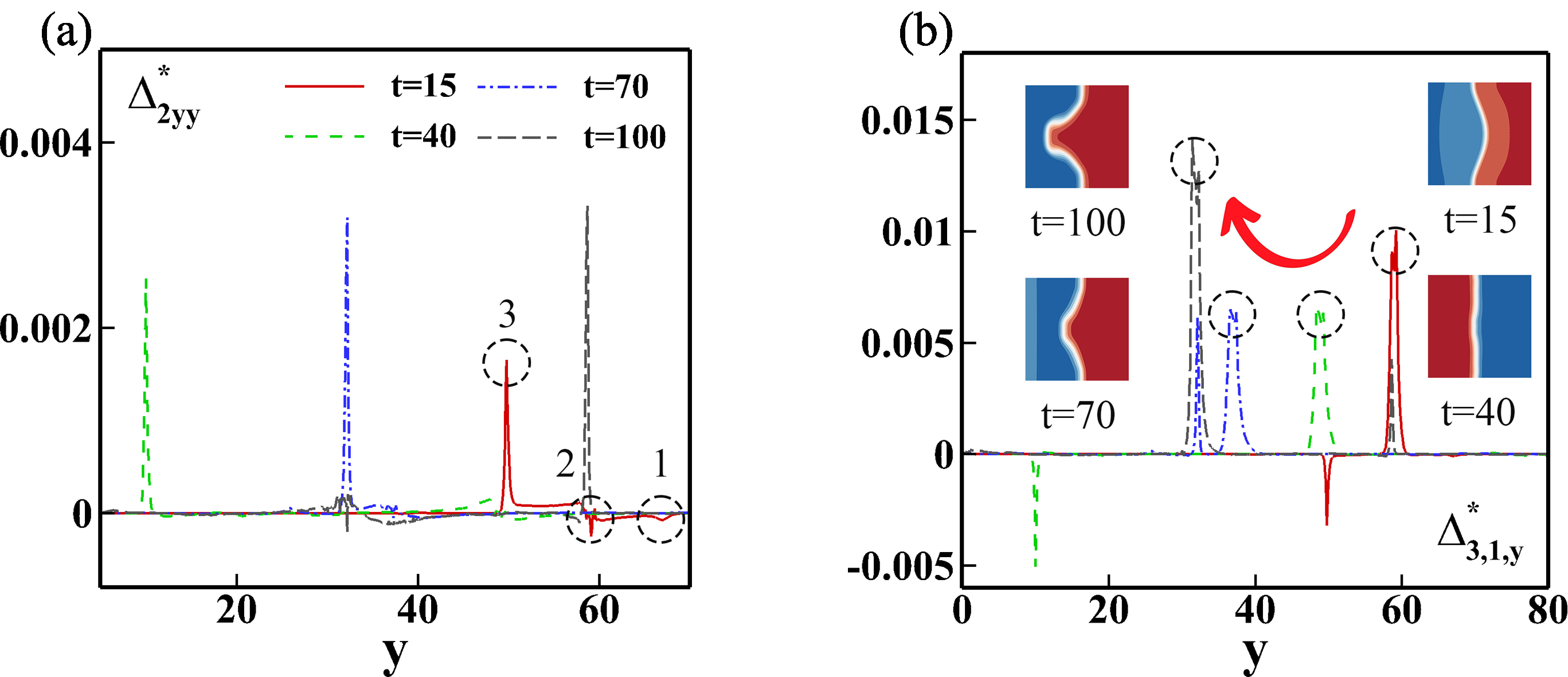}}
  \caption{Distribution of  different TNE quantities along $N_{x}/2$ at $t=15$, $t=40$, $t=70$ and $t=100$. (a)$\Delta_{2yy}^{*}$, (b)$\Delta_{3,1y}^{*}$.}
\label{fig:9}
\end{figure*}

In order to quantitatively investigate the evolution of non-equilibrium effects with time, the axe $x=N_{x}/2$ is selected and observed to see how the distribution of TNE quantities on the axes evolve over time. 
Figure~\ref{fig:9}(a) shows the distribution of the component $\Delta_{2yy}^{*}$ of NOMF $\boldsymbol{\Delta}_2^{*}$. It can be seen from the red line that $\Delta _{2yy}^{*}$ mainly distributes in three areas marked with "1", "2", and "3" from right to left.
Those three areas exactly correspond to rarefaction, material interface, and transmission shock wave. Besides, the strength of $\Delta _{2yy}^{*}$ near the material interface is stronger than rarefaction but weaker than transmission shock wave. 
The intensity of $\Delta _{2yy}^{*}$ near interface gradually decreases, but the intensity of $\Delta _{2yy}^{*}$  near transmission shock wave gradually increases.
Figure~\ref{fig:9}(b) shows the distribution of the component $\Delta_{3,1y}^{*}$ of NOEF $\boldsymbol{\Delta}_{3,1}^{*}$. It can be seen from the red line that the strength of $\Delta_{3,1y}^{*}$ near the material interface is much stronger than that near the transmission shock wave. After the reflection of the transmission shock wave, the direction of $\Delta_{3,1y}^{*}$ reverses, which is different from $\Delta_{2yy}^{*}$. 
By further observing the distribution of $\Delta_{3,1y}^{*}$ near the material interface, it is found that the intensity of  $\Delta_{3,1y}^{*}$ first decreases and then increases. 
The decrease of $\Delta_{3,1y}^{*}$ is mainly due to the increase of mixing area near the interface, where the gradients of physical quantities decrease.
The increase of $\Delta_{3,1y}^{*}$ is due to the passing of secondary shock, where the energy is transported from the shock to the material interface.
Besides, there exists a double-peak structure near the material interface.
In this case, the $\Delta_{3,1y}^{*}$ first gets its maximum value at the right peak and then the left one. The shift of the maximum peak depends on the direction of the shock wave.
\subsection{RMI with magnetic field}
The magnetic field can suppress the evolution of RMI through transporting the barometrically generated vorticity away from the interface \citep{samtaney2003,sano2013}.
In this section, the effects of magnetic fields on the evolution of RMI (including HNE and TNE effects) are analyzed. 
The initial magnetic field is set on the $y$ direction.
Table \ref{tab:table2} shows the parameters setting in different cases, where $\beta = 2 p_0 / B_0^2 $ is the nondimensional strength of the magnetic field. \cite{samtaney2003}
The other parameters are consistent with those in Sec.~\ref{settings}.
\begin{table}
\caption{\label{tab:table2}Settings for RMI with different initial applied magnetic fields. }
\begin{ruledtabular}
\begin{tabular}{lcclcc}
Cases & $B_{0}$ & $\beta$ & Cases & $B_{0}$ & $\beta$  \\
\hline
Case I   & 0.01 & $2.80 \times 10^4$&Case VI   & 0.10&$280.0$\\
Case II  & 0.02 & $7.00 \times 10^3$&Case VII  & 0.15&$124.4$\\
Case III & 0.03 & $3.11 \times 10^3$&Case VIII & 0.20&$70.0$\\
Case IV  & 0.04 &$1.75 \times 10^3$ &Case IX   & 0.25&$44.8$\\
Case V   & 0.05 & $1.12 \times 10^3$&Case X    & 0.30&$31.1$\\
\end{tabular}
\end{ruledtabular}
\end{table}
\subsubsection{HNE characteristics}
\begin{figure*}
  \centerline{\includegraphics[width=\linewidth]{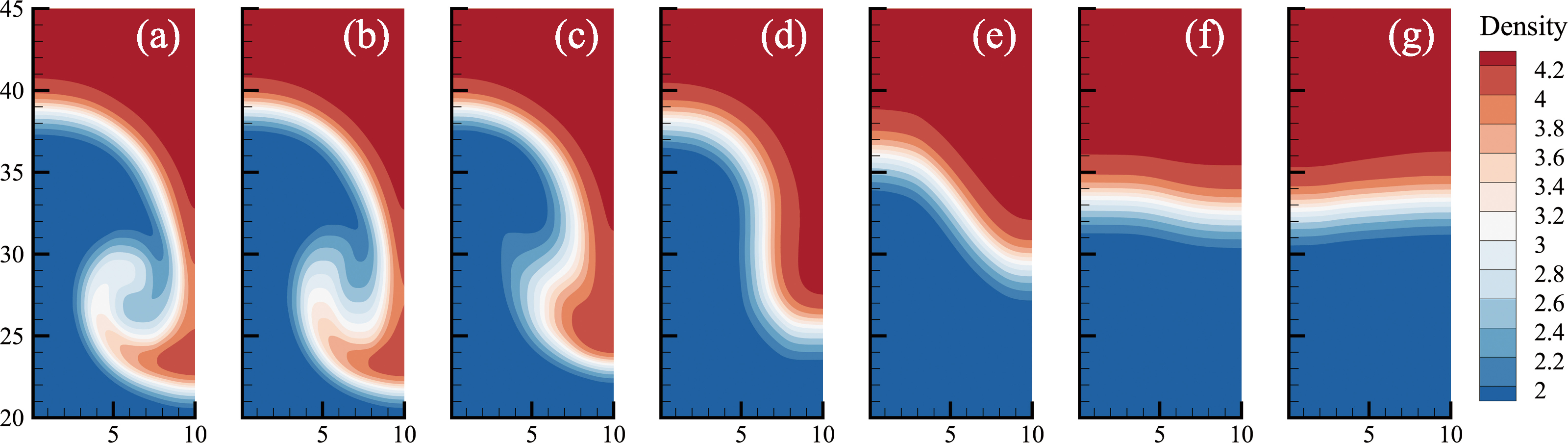}}
  \caption{Density contours with different applied magnetic fields at $t=250$. (a)$B_{0}=0.00$, (b)$B_{0}=0.01$, (c)$B_{0}=0.03$, (d)$B_{0}=0.05$, (e)$B_{0}=0.10$, (f)$B_{0}=0.20$, (g)$B_{0}=0.30$}
\label{fig:10}
\end{figure*}
Figure~\ref{fig:10} shows the density contours of different initial magnetic fields ranging from $0.00$ to $0.30$ at $t=250$. It is found that the evolution of the interface is gradually suppressed with the increase of magnetic fields. 
For the case of $B_0=0.01$, the Kelvin-Helmholtz instability (KHI) is still developed and observed, forming the "mushroom-like" structure.
For the case of $B_0=0.03$ and $B_0=0.05$, the KHI degenerates and the amplitude of the interface decreases.
The interface still reverses for the cases of $B_0 \le 0.20$ (figures~\ref{fig:10} (a)-(f)), but no longer reverses for the case of $B_0=0.30$ ( fig.~\ref{fig:10} (g)).
This indicates that there exists a critical magnetic field $B_{0,C}$, under which the amplitude of interface can be suppressed to $0$. 
\begin{figure}
  \centerline{\includegraphics[width=8cm]{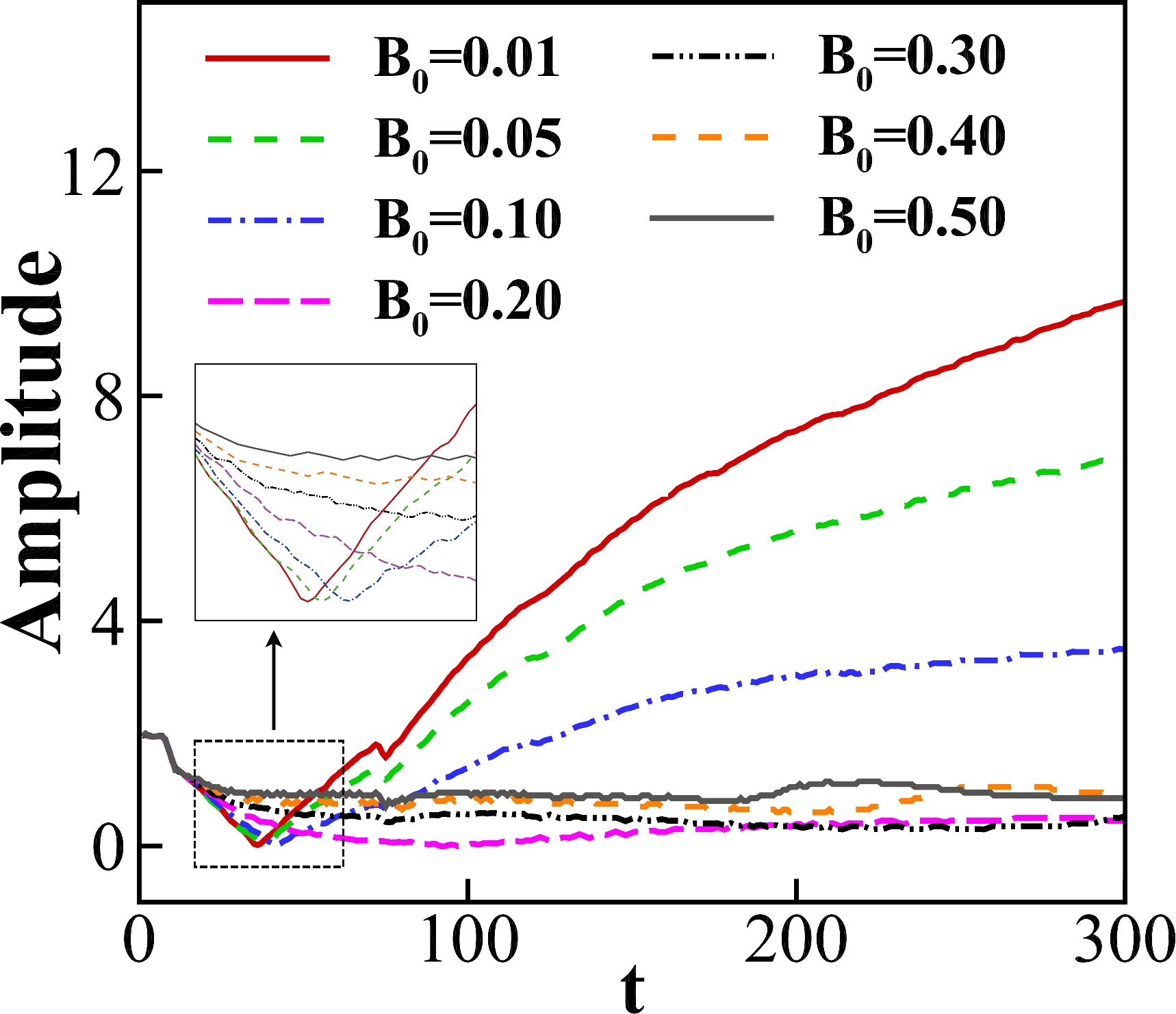}}
  \caption{Evolution of amplitude of RMI with different initial applied magnetic fields from $t=0$ to $t=200$.}
\label{fig:12}
\end{figure}

Figure~\ref{fig:12} further shows the evolution of interface amplitude with time.
Obviously, the magnetic fields delay the interface inversion time.
For the cases of $B_0 \le 0.05$, the magnetic fields have little effect on the flow field before the interface inversion but can significantly inhibit the development of interface amplitude after the interface inversion.
With the increase of the magnetic fields, the influence of the secondary shock on the perturbation amplitude is weakened. 
The reason is that the magnetic field inhibits the development of the interface, which reduces the perturbation amplitude when the secondary shock passes through the interface. 
Thus, the induced vorticity reduces.
With the increase of magnetic fields, a critical situation appears where the perturbation is nearly inhibited to $0$.
As the magnetic field further increases, the interface no longer reverses, and the vorticity induced by secondary shock will prevent the development of the interface.
When the magnetic field exceeds the critical magnetic field, the stronger the magnetic field, the closer the interface shape is to the initial state. 
\begin{figure*}
  \centerline{\includegraphics[width=17cm]{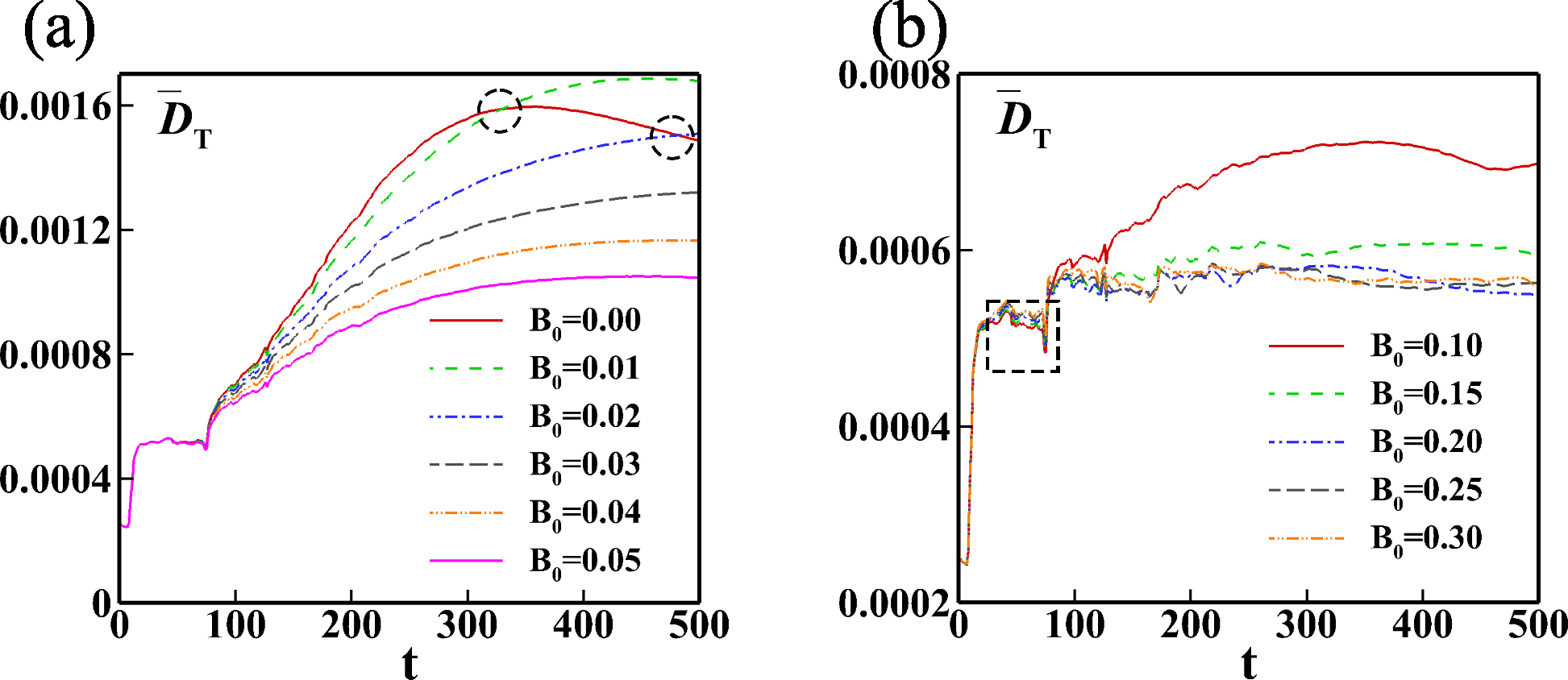}}
  \caption{Evolution of global average TNE effects with different initial applied magnetic fields from $t=0$ to $t=500$.}
\label{fig:13}
\end{figure*}
\begin{figure*}
  \centerline{\includegraphics[width=17cm]{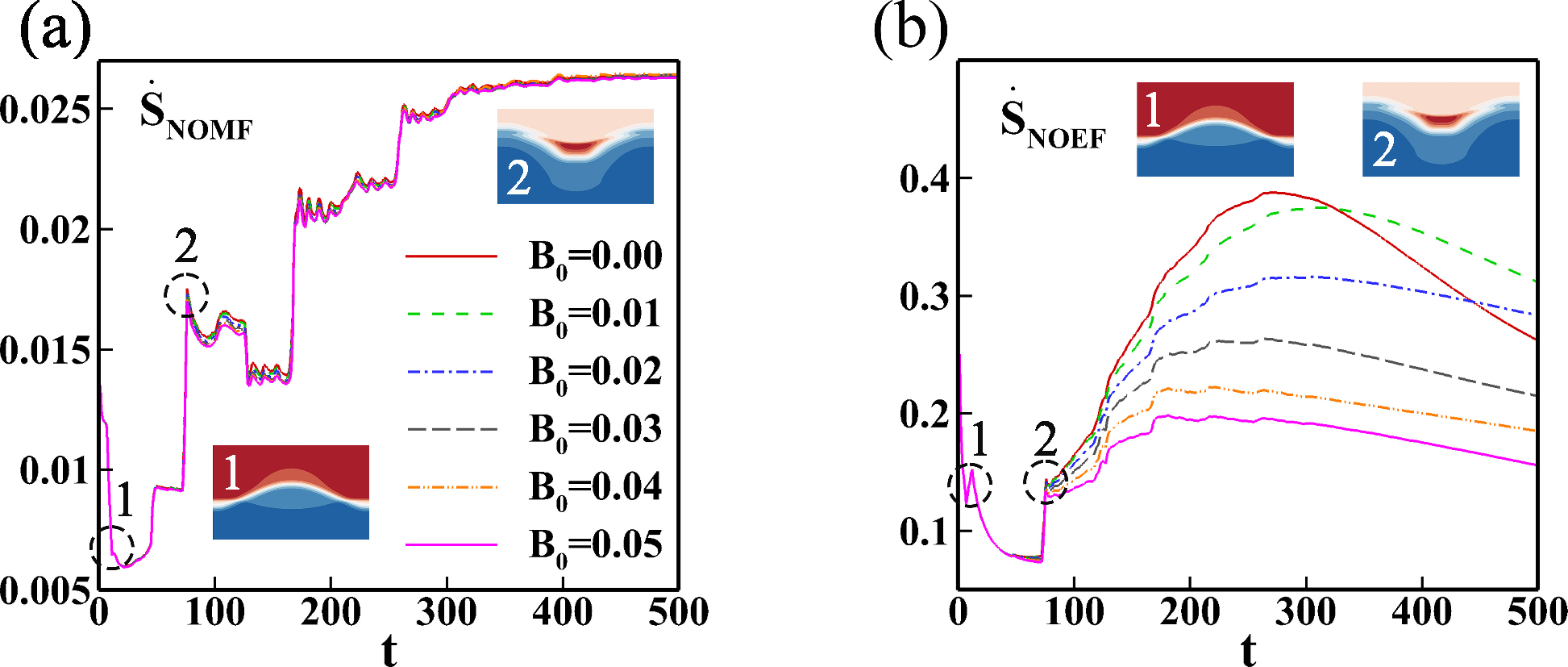}}
  \caption{Evolution of entropy production rates from $t=0$ to $t=500$ with magnetic fields ranging from $0.01$ to $0.05$.}
\label{fig:14}
\end{figure*}
\begin{figure*}
  \centerline{\includegraphics[width=17cm]{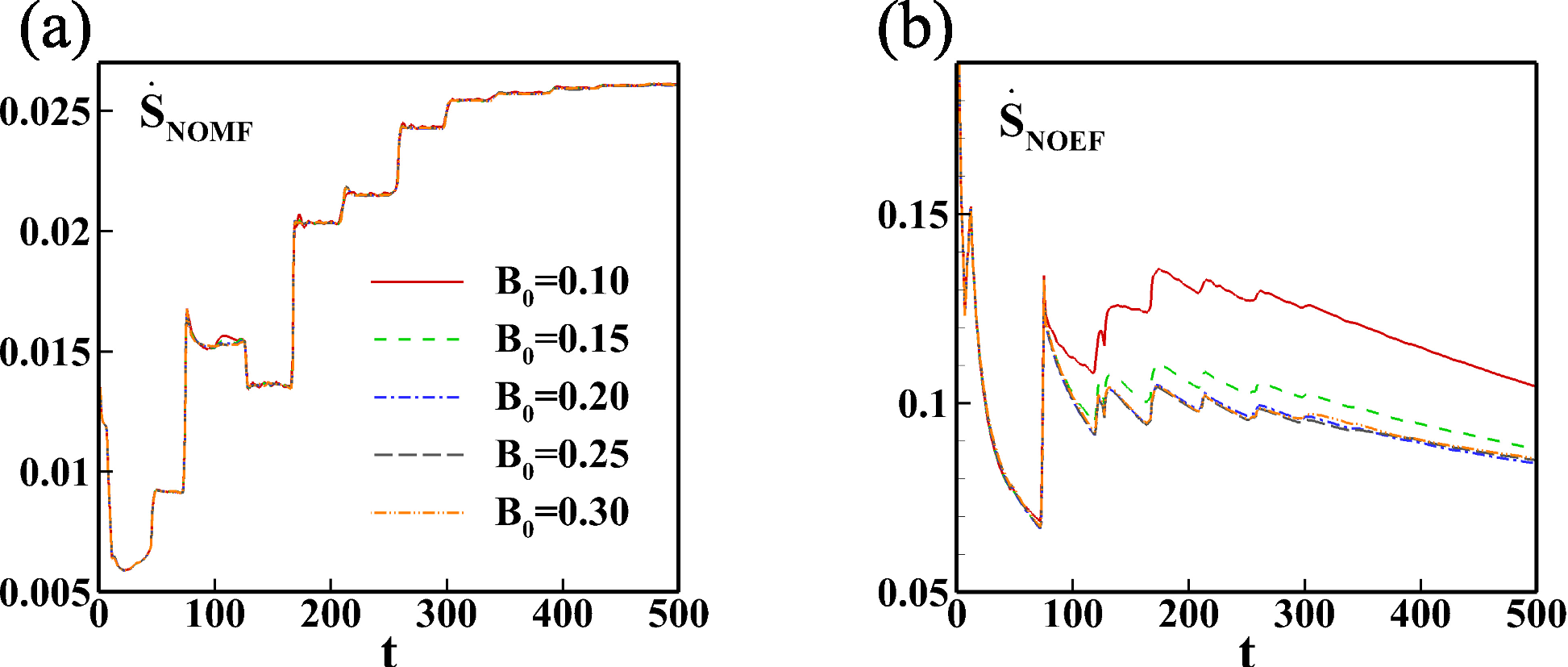}}
  \caption{Evolution of entropy production rates from $t=0$ to $t=500$ with magnetic fields ranging from $0.10$ to $0.30$.}
\label{fig:15}
\end{figure*}
\subsubsection{TNE effects}
Figure~\ref{fig:13}(a) shows the global average TNE strength $\bar D_T$ with different initial magnetic fields ranging from $0.00$ to $0.05$.
It is observed that, before the secondary shock, the increase of magnetic fields shows little effect to $\bar D_T$.
After the secondary shock, $\bar D_T$ quickly increases with the evolution of the interface, and the magnetic fields show great effects for suppressing $\bar D_T$. 
However, it is found that $\bar D_T$ for the case of $B_0=0.01$ and $B_0=0.02$ are greater than that of $B_0=0.00$ after the time point marked with a circle.
This is because the magnetic fields suppress the shear effects around the interface, thus delaying the time KHI arises and decreasing the strength of KHI.
The emergence and development of KHI can enhance mixing and reduce the gradients of physical quantities, which is helpful for reducing $\bar D_T$.
The inflection point of TNE intensity can be regarded as the criterion to judge whether KHI is fully developed. Before the inflection point, the development of perturbation amplitude dominates and the TNE is enhanced.
After the inflection point, the development of KHI results in the enhancement of dissipation effect and the decrease of TNE intensity.

Figure~\ref{fig:13}(b) shows the global average TNE strength $\bar D_T$ with different initial magnetic fields ranging from $0.10$ to $0.30$.
It is found that the influence of magnetic field for TNE could be divided into two stages. 
Before the interface inversion, as the strength of the magnetic field increases, the TNE strength slightly increases. 
After the interface inversion, the TNE strength decreases significantly with the increase of magnetic field strength. 
In fact, the magnetic field consistently inhibits the interface development during the RMI evolution. 
Before the interface inversion, the magnetic field inhibits the interface inversion, indirectly enhancing the physical quantity gradient near the interface, resulting in the enhancement of TNE intensity. 
After the interface inversion, the magnetic field inhibits the shear effect near the interface, as has been explained before.
From Fig.~\ref{fig:13}(b), it can also be found that after secondary, $\bar D_T$ keeps decreasing with the increase of magnetic fields, and there exists a critical magnetic field, above which $\bar D_T$ no longer decreases.
Thus, the magnetic field intensity $B_{0,C}$, corresponding to the minimum value $\bar D_{T,min}$ of global average TNE intensity just after the re-shock stage, can be used as the critical magnetic field intensity to inhibit the development of RMI interface.
\subsubsection{Entropy production}
Figure~\ref{fig:14} shows the two parts of entropy production rate ${{\dot S}_{NOMF}}$ and ${{\dot S}_{NOEF}}$ of the global system with magnetic fields range from $0.01$ to $0.05$. 
From Fig.~\ref{fig:14}(a) we can find that before the shock wave contacts the perturbed interface, the rate sharply decreases. 
Then, after the shock wave contacts the solid wall and passes through the interface again, the rate appears two jumps. 
After re-shock, the rate first decreases and then increases to the maximum. 
With the increase of magnetic fields, the values of ${{\dot S}_{NOMF}}$ are basically the same.
From Fig.~\ref{fig:14}(b), it is observed that when the shock first contacts the interface, the entropy production rate ${{\dot S}_{NOEF}}$ suddenly increases (point ‘1'). 
It means that the shock wave enhances physical quantity gradients near the interface. 
When the transmitted shock wave passes the interface again, as shown by point ‘2', the rate increases greatly, which is also observed from ${{\dot S}_{NOMF}}$.
After re-shock, the rate first gradually increases and then decreases.
When the initial magnetic field strength gradually increases, the  ${{\dot S}_{NOEF}}$ also shows no difference before re-shock but greatly reduces after re-shock.
In general, the ${{\dot S}_{NOEF}}$ contributes more to entropy increase than ${{\dot S}_{NOMF}}$, but the strength of ${{\dot S}_{NOEF}}$ could be greatly inhibited by adding magnetic field.

Figure~\ref{fig:15} shows the two parts of entropy production rate ${{\dot S}_{NOMF}}$ and ${{\dot S}_{NOEF}}$ of the global system with magnetic fields range from $0.10$ to $0.30$. 
It can observed that the fluctuations in the ${{\dot S}_{NOMF}}$ curve have been smoothed due to the increase of magnetic fields (Fig.~\ref{fig:14}(a) and Fig.~\ref{fig:15}(a)). 
Differ from ${{\dot S}_{NOMF}}$, the evolution of ${{\dot S}_{NOEF}}$ is strongly affected by magnetic fields. 
With the increase of magnetic fields, the evolution of ${{\dot S}_{NOEF}}$ after re-shock is significantly inhibited. 
It can be deduced that there exists a critical magnetic field, under which the ${{\dot S}_{NOEF}}$ no longer decreases.
\section{Discussion and conclusion}
\label{Conclusion}
When the particle collision frequency is sufficiently high, the kinetic behaviors can be described by the reduced hydrodynamic equations. When the particle collision frequency is negligibly small, the kinetic behaviors can be described by the reduced Vlasov equation where the collision effect is completely ignored. 
The situation between the two extreme cases is generally difficult to treat with and consequently poorly understood, which is responsible for the fact that the kinetic effects caused by particle collision in ICF are still far from clear understanding although they have a potential  impact on ICF ignition. 

To solve the above problems, the preferred method/model needs to simultaneously have two physical functions.
First, it can construct a physical model to recover the HNE and TNE behaviors we aim to investigate before simulation. 
Second, it can present methods for checking the non-equilibrium state, describing and analyzing the resulting effects from massive data after simulation. 
In the DBM, for the first physical function, the model equations are composed of a discrete Boltzmann equation coupled with a magnetic induction equation. For the second physical function, DBM intrinsically includes some schemes for studying the TNE behaviors. The most fundamental one is to use the non-conserved kinetic moments of $(f-f^{eq})$ to check the TNE state and describe the TNE behaviors. The phase space description method based on the non-conserved kinetic moments of $(f-f^{eq})$ presents an intuitive geometrical correspondence for the complicated TNE state and behaviors, which is important for deep investigation and clear understanding. 
The first function is verified by recovering HNE behaviors of a number of typical benchmark problems including sod shock tube, thermal Couette flow, and OT vortex problem. Besides, the most relevant TNE behaviors of the OT vortex problem are investigated for the first time.  
As a further application, the kinetic study on the RMI system with initial horizontal magnetic fields is performed.  

Physical findings are as below:
(i) Generally, the entropy production rate is in positive correlation to the difficulty of compression. In the process of OT evolution, the difficulty of compression is divided into stages: in the first stage, the entropy production rate and compression difficulty increase with time; while in the second stage, the entropy production rate and compression difficulty decrease with time. 
(ii) In the case without an external magnetic field, the NOMF gets its maximum value near the shock front, while the NOEF gets its maximum value near the perturbed interface.  During the perturbed interface inversion process of RMI, the NOEF along the central axis shows  two peak values near the perturbed interface, and the impact of the shock wave significantly enhances the NOEF.
(iii) In the cases with external magnetic fields, the magnetic field shows inhibitory effects on the evolution of RMI. Specifically, the magnetic field has a pronounced inhibitory effect on the nonlinear stage, especially on the generation of KHI. Besides, there exists a critical magnetic field under which the amplitude of the interface can be suppressed to $0$.
(iv) Before the interface inversion, the magnetic field indirectly enhances the TNE intensity by suppressing the interface inversion. After the interface inversion, the magnetic field significantly suppresses the TNE intensity by inhibiting the further development of the perturbed interface. 
(v) In terms of entropy production rate, the magnetic field has a pronounced inhibitory effect on the entropy production rate caused by heat conduction.
However, the effects of magnetic field on the entropy production rate caused by viscosity is small.

Potential applications of the above physical findings are as follows: 
(i) Based on features of NOMF and NOEF, shock front and perturbed interface during RMI evolution can be physically captured in the numerical experiments.
(ii) The magnetic field intensity $B_{0,C}$, corresponding to the minimum value $\bar D_{T,min}$ of global average TNE intensity just after the re-shock stage, can be used as the critical magnetic field intensity to inhibit the development of RMI interface.
(iii) From the perspective of entropy production rate, the magnetic field intensity $B_{0,C}$, corresponding to the minimum value of entropy production rate caused by heat conduction just after the re-shock stage, plays the same role.

The DBM constructed in this article is still relatively simple.
In the future, we will further construct DBM with more physical functions. Besides, the DBM will be used to investigate more complex plasma systems.
Possible research directions are as follows:
(i) Two-fluid DBM for plasma. Based on the two-fluid magnetic fluid model, \citet{bond2017richtmyer} found that charge separation will cause the electron interface to rapidly develop fine-scale structures. 
These structures challenge the reasonability of the single-fluid model, so it is necessary to develop the ion-electron two-fluid model.
(ii) New DBM that considers finite electric conductivity. Physically, the effects of viscous stress, heat conduction, and electrical conductivity are all caused by particle collisions. The model in this paper considers a simplified case where the electrical conductivity is assumed to be infinity. Thus, DBM that considers finite electric conductivity deserves to be further constructed.
(iii) The coupling relationship between TNE quantities and macroscopic quantities. \citet{rinderknecht2018} have pointed out that kinetic physics caused by particle collisions may affect ICF ignition. The main difficulty they face in numerical simulation is that the influence of kinetic physics does not necessarily produce independent phenomena that can be observed. It tends to have a weak influence on macroscopic quantities. The TNE quantity given by DBM has the potential for investigating the kinetic physics in plasma, so the coupling relationship between TNE quantities and macroscopic quantities deserves further study.
(iv) RMI under the combined action of re-shock and magnetic field. 
It is known that both re-shock and external magnetic fields can inhibit the development of RMI.
However, their combined effect has not been fully explored. 
This article presents several physical findings, which will be used to investigate the mechanism when re-shock and magnetic fields act together.
\begin{acknowledgments}
The authors sincerely thank Lingxiao Li for his helpful suggestion on choosing the discrete format. The authors thank Yudong Zhang, Chuandong Lin, Ge Zhang, Dejia Zhang, Yiming Shan, Jie Chen, Hanwei Li, and Yingqi Jia for the helpful discussion on DBM. The authors thank Song Bai, Fuwen Liang, Feng Tian, Zihao He, Mingqing Nie, and Zhengxi Zhu for the helpful discussion on the results analysis.
This work was supported by the National Natural Science Foundation of China (grant numbers 52202460, 12172061 and 11975053), the Foundation of National Key Laboratory of Shock Wave and Detonation Physics (grant numbers JCKYS2023212003), the National Key R D Program of China (grant numbers 2020YFC2201100, 2021YFC2202804, 2022YFB3403504), the Natural Science Foundation of Shandong Province (grant numbers ZR2020MA061), the opening project of State Key Laboratory of Explosion Science and Technology (Beijing Institute of Technology) (grant numbers KFJJ2023-02M).
\end{acknowledgments}

\section*{Data Availability Statement}

The data that support the findings of this study are available from the corresponding authors upon reasonable request.

\appendix
\section{Kinetic moments used to characterize the first-order and second-order TNE effects}
\label{appA}
Here, the kinetic moments used to characterize the first-order and second-order TNE effects are given as below,
\begin{eqnarray}
M_0^{eq} = \sum\limits_i {f_i^{eq}}  = \rho, \label{eq:A1}
\end{eqnarray}
\begin{eqnarray}
M_{1,x}^{eq} = \sum\limits_i {f_i^{eq}{v_{ix}}}  = \rho {u_x}, \label{eq:A2}
\\
M_{1,y}^{eq} = \sum\limits_i {f_i^{eq}{v_{iy}}}  = \rho {u_y}, \label{eq:A3}
\end{eqnarray}
\begin{eqnarray}
M_{2,0}^{eq} &=& \sum\limits_i {f_i^{eq}\left( {v_{ix}^2 + v_{iy}^2 + \eta _i^2} \right)}  \nonumber\\
&=& \rho \left[ {\left( {n + 2} \right)RT + u_x^2 + u_y^2} \right], \label{eq:A4}
\\
M_{2,xy}^{eq} &=& \sum\limits_i {f_i^{eq}{v_{ix}}{v_{iy}}}  = \rho {u_x}{u_y}, \label{eq:A5}
\\
M_{2,xx}^{eq} &=& \sum\limits_i {f_i^{eq}{v_{ix}}{v_{ix}}}  = \rho \left( {RT + u_x^2} \right), \label{eq:A6}
\\
M_{2,yy}^{eq} &=& \sum\limits_i {f_i^{eq}{v_{iy}}{v_{iy}}}  = \rho \left( {RT + u_y^2} \right), \label{eq:A7}
\end{eqnarray}
\begin{eqnarray}
M_{3,1,x}^{eq} &=& \sum\limits_i {f_i^{eq}\left( {v_{ix}^2 + v_{iy}^2 + \eta _i^2} \right){v_{ix}}}  \nonumber\\
&=& \rho {u_x}\left[ {\left( { n + 4} \right)RT + u_x^2 + u_y^2} \right], \label{eq:A8}
\\
M_{3,1,y}^{eq} &=& \sum\limits_i {f_i^{eq}\left( {v_{ix}^2 + v_{iy}^2 + \eta _i^2} \right){v_{iy}}}   \nonumber\\
&=& \rho {u_y}\left[ {\left( { n + 4} \right)RT + u_x^2 + u_y^2} \right], \label{eq:A9}
\end{eqnarray}
\begin{eqnarray}
M_{3,xxx}^{eq} &=& \sum\limits_i {f_i^{eq}v_{ix}^3}  = \rho {u_x}\left( {3RT + u_x^2} \right), \label{eq:A10}
\\
M_{3,yyy}^{eq} &=& \sum\limits_i {f_i^{eq}v_{iy}^3}  = \rho {u_y}\left( {3RT + u_y^2} \right), \label{eq:A11}
\\
M_{3,xxy}^{eq} &=& \sum\limits_i {f_i^{eq}v_{ix}^2{v_{iy}}}  = \rho {u_y}\left( {RT + u_x^2} \right), \label{eq:A12}
\\
M_{3,xyy}^{eq} &=& \sum\limits_i {f_i^{eq}{v_{ix}}v_{iy}^2}  = \rho {u_x}\left( {RT + u_y^2} \right), \label{eq:A13}
\end{eqnarray}
\begin{eqnarray}
M_{4,2,xx}^{eq}  &=& \sum\limits_i {f_i^{eq}\left( {v_{ix}^2 + v_{iy}^2 + \eta _i^2} \right)v_{ix}^2}  \nonumber\\
 &=&  \rho [(n+4) R^2 T^2+R T \left((n+7) u_x^2+u_y^2\right) \nonumber\\
&&+ u_x^2 \left(u_x^2+u_y^2\right)], \label{eq:A14}
\\
M_{4,2,yy}^{eq}  &=&  \sum\limits_i {f_i^{eq}\left( {v_{ix}^2 + v_{iy}^2 + \eta _i^2} \right)v_{iy}^2}  \nonumber\\
 &=&  \rho [(n+4) R^2 T^2+R T \left((n+7) u_y^2+u_x^2\right)\nonumber\\
&&+u_y^2 \left(u_x^2+u_y^2\right)], \label{eq:A15}
\\
M_{4,2,xy}^{eq} &=& \sum\limits_i {f_i^{eq}\left( {v_{ix}^2 + v_{iy}^2 + \eta _i^2} \right){v_{ix}}} {v_{iy}} \nonumber\\
&=& \rho {u_x}{u_y}\left[ {\left( { n + 6} \right)RT + u_x^2 + u_y^2} \right], \label{eq:A16}
\end{eqnarray}
\begin{eqnarray}
M_{4,xxxx}^{eq} &=& \sum\limits_i {f_i^{eq}v_{ix}^4}  = \rho \left(3 R^2 T^2+6 R T u_x^2  + u_x^4\right), \label{eq:A17}
\\
M_{4,yyyy}^{eq} &=& \sum\limits_i {f_i^{eq}v_{iy}^4}  = \rho \left(3 R^2 T^2+6 R T u_y^2  + u_y^4\right), \label{eq:A18}
\\
M_{4,xxxy}^{eq} &=& \sum\limits_i {f_i^{eq}v_{ix}^3v_{iy}}  = \rho u_x u_y \left(3 R T+u_x^2\right), \label{eq:A19}
\\
M_{4,xyyy}^{eq} &=& \sum\limits_i {f_i^{eq}v_{ix}v_{iy}^3}  = \rho u_x  u_y \left(3 R T+u_y^2\right), \label{eq:A20}
\\
M_{4,xxyy}^{eq} &=& \sum\limits_i {f_i^{eq}v_{ix}^2v_{iy}^2}  = \rho  \left(R T+u_x^2\right) \left(R T+u_y^2\right), \label{eq:A21}
\end{eqnarray}
\begin{eqnarray}
M_{5,3,xxx}^{eq} &=& \sum\limits_i {f_i^{eq}\left( {v_{ix}^2 + v_{iy}^2 + \eta _i^2} \right)v_{ix}^3} \nonumber\\
   &=&  \rho u_x [3 (n+6) R^2 T^2+R T \left((n+11) u_x^2+3 u_y^2\right) \nonumber\\
  &&+u_x^2 \left(u_x^2+u_y^2\right)],  \label{eq:A22}
\\
M_{5,3,yyy}^{eq}  &=&  \sum\limits_i {f_i^{eq}\left( {v_{ix}^2 + v_{iy}^2 + \eta _i^2} \right) v_{iy}^3} \nonumber\\ 
 &=&  \rho u_y [3 (n+6) R^2 T^2+R T \left((n+11) u_y^2+3 u_x^2\right) \nonumber\\
&&+u_y^2 \left(u_x^2+u_y^2\right)], \label{eq:A23}
\\
M_{5,3,xxy}^{eq}  &=&  \sum\limits_i {f_i^{eq}\left( {v_{ix}^2 + v_{iy}^2 + \eta _i^2} \right)v_{ix}^2 v_{iy}} \nonumber\\ 
 &=&  \rho u_y [(n+6) R^2 T^2+R T \left((n+9) u_x^2+u_y^2\right) \nonumber\\
&&+u_x^2 \left(u_x^2+u_y^2\right)], \label{eq:A24}
\\
M_{5,3,xyy}^{eq}  &=& \sum\limits_i {f_i^{eq}\left( {v_{ix}^2 + v_{iy}^2 + \eta _i^2} \right)v_{ix} v_{iy}^2} \nonumber\\ 
 &=&  \rho u_x [(n+6) R^2 T^2+R T \left((n+9) u_y^2+u_x^2\right) \nonumber\\
&&+u_y^2 \left(u_x^2+u_y^2\right)],  \label{eq:A25}
\end{eqnarray}

where $p=\rho RT$ obeys the ideal gas law. 
\section{Chapman-Enskog multi-scale analysis}
\label{appB}
Through constructing kinetic moments ${{\bf{M}}_0}$, ${{\bf{M}}_1}$, and ${{\bf{M}}_{2,0}}$ on both sides of Eq~\eqref{eq:1}, the generalized hydrodynamic equations\footnote{where the viscous stress and heat conduction may contain not only the contribution of $f^{(1)}$  , among which the NS viscous stress and NS heat conduction are the simplest cases} could be obtained as follows,
\begin{equation}
\frac{{\partial \rho }}{{\partial t}} + \nabla  \cdot \left( {\rho {\bf{u}}} \right) = 0,
\label{eq:B1}
\end{equation}
\begin{equation}
\frac{{\partial \rho {\bf{u}}}}{{\partial t}} + \nabla  \cdot \left( {\rho {\bf{uu}} + {{\boldsymbol{\Delta }}_2}} \right) = \rho \bf{a},
\label{eq:B2}
\end{equation}
\begin{equation}
\frac{{\partial {E_T}}}{{\partial t}} + \nabla  \cdot \left[ {{E_T}{\bf{u}} + {{\boldsymbol{\Delta }}_{3,1}}} \right] = \rho \bf{a} \cdot \bf{u},
\label{eq:B3}
\end{equation}
where ${E_T} = \rho e + \rho {u^2}/2$ is the total energy, and $e$ is the energy density. ${{\boldsymbol{\Delta }}_{2}}$ and ${{\boldsymbol{\Delta }}_{3,1}}$ are two non-equilibrium quantities defined by Eq.~\eqref{eq:10}.
The relationship between ${{\boldsymbol{\Delta }}_{2}}$ (${{\boldsymbol{\Delta }}_{3,1}}$) and ${\boldsymbol{\Delta }}_2^*$  (${\boldsymbol{\Delta }}_{3,1}^*$) are given in reference \citep{gan_xu2022} as follows,
\begin{equation}
{{\boldsymbol{\Delta }}_2} = {\boldsymbol{\Delta }}_2^*,
\label{eq:B6}
\end{equation}
\begin{equation}
{{\boldsymbol{\Delta }}_{3,1}} = {\boldsymbol{\Delta }}_{3,1}^* + {\boldsymbol{\Delta }}_2^* \cdot {\bf{u}}.
\label{eq:B7}
\end{equation}

Here, ${\boldsymbol{\Delta }}_2^*$ is defined as non-organized momentum flux (NOMF) , and ${\boldsymbol{\Delta }}_{3,1}^*$ is defined as non-organized energy flux (NOEF). 
Compared with the constitutive of viscous stress and heat flux in NS, Burnett and Super-Burnett equations, ${\boldsymbol{\Delta }}_2^*$ and ${\boldsymbol{\Delta }}_{3,1}^*$ contain the most complete constitutive information. 
Thus, ${\boldsymbol{\Delta }}_2^*$ and ${\boldsymbol{\Delta }}_{3,1}^*$ are also referred to as generalized viscous stress and heat flux.
However, the specific forms of ${\boldsymbol{\Delta }}_2^*$ and ${\boldsymbol{\Delta }}_{3,1}^*$ are basically unknown. 
To get the specific forms of ${\boldsymbol{\Delta }}_2^*$ and ${\boldsymbol{\Delta }}_{3,1}^*$, we need to determine the order of $f$ to be retained (or the number of kinetic moments to be preserved) through CE multi-scale analysis. 
The form of CE multi-scale analysis is as follows,
\begin{equation}
f = {f^{(0)}} + Kn{f^{(1)}} + K{n^2}{f^{(2)}} + K{n^3}{f^{(3)}} +  \cdots ,
\label{eq:B8}
\end{equation}
\begin{equation}
\frac{\partial }{{\partial t}} = Kn\frac{\partial }{{\partial {t_1}}} + K{n^2}\frac{\partial }{{\partial {t_2}}} +  \cdots ,
\label{eq:B9}
\end{equation}
\begin{equation}
\frac{\partial }{{\partial {\bf{r}}}} = Kn\frac{\partial }{{\partial {{\bf{r}}_1}}},
\label{eq:B10}
\end{equation}
where ${f^{(0)}}={f^{eq}}$. 
Here, the zero-order expansions of the time and space derivatives $\partial /\partial {t_0}$ and $\partial /\partial {\bf{r}_0}$ are neglected, for the subscript $0$ represents the scale of system and the internal change of system cannot be observed through system scale. 
Besides, the time is expanded to the second-order and the space is expanded to the first-order. 
Through this way, the derived equations are exactly the hydrodynamic equations currently used such as NS and Burnett, etc.
Theoretically, it is also optional to expand time to the first-order and expand space to the second-order. 
As long as the derivation process is correct, the obtained hydrodynamic equations are also correct.
Substituting Eqs.~(\ref{eq:B8})-(\ref{eq:B8}) into Eq.~(\ref{eq:1}),  and retaining the same order terms of $Kn$ numbers, we get,
\begin{equation}
K{n^1}:\frac{{\partial {f^{(0)}}}}{{\partial {t_1}}} + \frac{{\partial \left( {{f^{(0)}}{\bf{v}}} \right)}}{{\partial {{\bf{r}}_1}}} =  - \frac{1}{\tau }{f^{(1)}},
\label{eq:B11}
\end{equation}
\begin{equation}
K{n^2}:\frac{{\partial {f^{(0)}}}}{{\partial {t_2}}} + \frac{{\partial {f^{(1)}}}}{{\partial {t_1}}} + \frac{{\partial \left( {{f^{(1)}}{\bf{v}}} \right)}}{{\partial {{\bf{r}}_1}}} =  - \frac{1}{\tau }{f^{(2)}}.
\label{eq:B12}
\end{equation}

By taking the ${{\bf{M}}_0}$, ${{\bf{M}}_1}$, and ${{\bf{M}}_{2,0}}$ moments simultaneously on both sides of Eq.~(\ref{eq:B11}) and use the relation ${f^{(1)}} =  - \tau \left[ {\partial {f^{(0)}}/\partial {t_1} + \partial ({f^{(0)}}{\bf{v}})/\partial {{\bf{r}}_1}} \right]$, the viscous stress and heat flux in NS level can be deduced after performing some mathematical transformations, as follows \cite{gan_xu2022},
\begin{equation}
{\boldsymbol{\Delta}}_{2, NS}^{*}={\boldsymbol{\Delta}}_2^{*(1)} = \int_{ - \infty }^\infty  {{f^{(1)}}{{\bf{v}}^*}{{\bf{v}}^*}d{\bf{v}}}  =  - \mu \left[ {\left( {\nabla {\bf{u}}} \right) + {{\left( {\nabla {\bf{u}}} \right)}^{\rm{T}}}} \right],
\label{eq:B13}
\end{equation}
\begin{equation}
{\boldsymbol{\Delta}}_{3,1, NS}^{*}={\boldsymbol{\Delta}}_{3,1}^{*(1)} = \int_{ - \infty }^\infty  {\frac{1}{2}{f^{(1)}}{{\bf{v}}^*} \cdot {{\bf{v}}^*}{{\bf{v}}^*}d{\bf{v}}}  =  - \kappa \nabla T,
\label{eq:B14}
\end{equation}
where $\mu = \tau \rho RT$ is the viscosity coefficient and $\kappa = 2\tau \rho RT$ is the heat conductivity. 
Similarly, the viscous stress ${\boldsymbol{\Delta}}_2^{*(2)}$ and heat flux ${\boldsymbol{\Delta}}_{3,1}^{*(2)}$ contributed by $f^{(2)}$ can be deduced from Eq.~(\ref{eq:B12}), and the viscous stress and heat flux in Burnett level are ${\boldsymbol{\Delta}}_{2, Burnett}^{*}={\boldsymbol{\Delta}}_2^{*(1)}+{\boldsymbol{\Delta}}_2^{*(2)}$ and 
${\boldsymbol{\Delta}}_{3,1, Burnett}^{*}={\boldsymbol{\Delta}}_{3,1}^{*(1)}+{\boldsymbol{\Delta}}_{3,1}^{*(2)}$, respectively.
It should be noted that, CE gives the dependence of  $(n+1)$-th order distribution function $f^{(n+1)}$ on the ${n}$-th order distribution function $f^{(n)}$, and finally gives the dependence of $f^{(n+1)}$ on $f^{(0)}$, where $f^{(0)}$ is the only one whose kinetic moments are known and can be relied upon in the modeling process. In the derivation of the first-order constitutive relations in NS, the highest-order kinetic moment used is ${\bf{M}}_{4,2}$. As for the derivation of the second-order constitutive relations in Burnett, the highest-order kinetic moment used is ${\bf{M}}_{5,3}$.
\section{Kinetic Macro Modeling and Kinetic Direct Modeling method}
\label{appC}
There exist two kinds of methods to obtain the macroscopic hydrodynamic equations. The first is the traditional macroscopic direct modeling method which is based on the continuum assumption and near equilibrium approximation. The second is the Kinetic Macro Modeling (KMM) method, that is, starting from kinetic equations, to derive the corresponding macroscopic hydrodynamic equations via CE multi-scale analysis. 
Due to different research ideas, KMM is divided into two categories. The first category follows the traditional modeling idea which concerns only the evolution equations of the conserved kinetic moments, that are, the density, momentum and energy. The second category realizes the insufficiency of the first category in capturing the system behaviors for the high $Kn$ cases and consequently derives also the evolution equations of the most relevant non-conserved kinetic moments \citep{chen2017}. 
For convenience of description, we refer the model equations derived from the second category of KMM to as extended hydrodynamic equations (EHE)\footnote{where the model equations include not only the evolution equations of density, momentum and energy, but also those of the most relevant non-conserved kinetic moments}.
Currently, most of the KMM studies belong to the first category. 

In contrast, the DBM is a Kinetic Direct Modeling method (KDM). Here `direct' means without needing to know the specific EHE. Because the current DBM is still working in the case where the CE theory is valid, the CE theory is the mathematical guarantee for rationality and effectiveness of DBM.
DBM is responsible for the following two aspects: (i) According to the discrete, non-equilibrium degree (described by Knudsen number), determine the system behavior needs to be grasped from what aspects, so as to determine which kinetic moments must be preserved in the process of model simplification, (ii) based on the non-conserved moments of $(f-f^{eq})$, present as many as possible schemes for the detection, description, presentation and analysis of the TNE state and effect.
For the first aspect, from the perspective of KMM, DBM determines the kinetic moments of $f^{eq}$ that need to be preserved according to the requirements to obtain the more accurate constitutive expressions; from the perspective of kinetic theory, DBM determines the kinetic moments of $f^{eq}$ that need to be preserved according to the requirements to obtain the more accurate distribution function expressions \citep{xu2022complex}.
Obviously, the second category of KMM is closest to DBM in physical function.

In addition to the equations for the conservation of mass, momentum, and energy, the extended hydrodynamic equations derived through KMM also includes equations describing the evolution of non-equilibrium quantities such as viscous stress and heat flux. 
For example, by integrating ${\bf{v}}^* {\bf{v}}^*$ and $\frac{1}{2}{{\bf{v}}^*} \cdot {{\bf{v}}^*}{{\bf{v}}^*}$ on both sides of Eq.~(\ref{eq:1}) and ignore the external force term, we obtain the following equation \citep{gan_xu2022},
\begin{eqnarray}
 && \frac{{\partial {\boldsymbol{\Delta}}_2^*}}{{\partial t}} + \frac{{\partial {\bf{M}}_2^*\left( {{f^{(0)}}} \right)}}{{\partial t}} + \\ \nonumber
 && \nabla  \cdot \left[ {{\bf{M}}_3^*\left( {{f^{(0)}}} \right) + {\bf{M}}_2^*\left( {{f^{(0)}}} \right){\bf{u}} + {\boldsymbol{\Delta}}_3^* + {\boldsymbol{\Delta}}_2^*{\bf{u}}} \right] =  - \frac{1}{\tau }{\boldsymbol{\Delta}}_2^*,
 \label{eq:C11}
\end{eqnarray}
\begin{eqnarray}
 && \frac{{\partial {\boldsymbol{\Delta}}_{3,1}^*}}{{\partial t}} + \frac{{\partial {\bf{M}}_{3,1}^*\left( {{f^{(0)}}} \right)}}{{\partial t}} + \\ \nonumber
 && \nabla  \cdot \left[ {{\bf{M}}_{4,2}^*\left( {{f^{(0)}}} \right) + {\bf{M}}_{3,1}^*\left( {{f^{(0)}}} \right){\bf{u}} + {\boldsymbol{\Delta}}_{4,2}^* + {\boldsymbol{\Delta}}_{3,1}^*{\bf{u}}} \right] \\ \nonumber
 &&=  - \frac{1}{\tau }{\boldsymbol{\Delta}}_{3,1}^*,
 \label{eq:C22}
\end{eqnarray}
where Eqs.~(\ref{eq:C11}) and (\ref{eq:C22}) describe the evolution of viscous stress and heat flux, respectively. 
Therefore, the obtaining of higher-order of TNE quantities such as ${\boldsymbol{\Delta}}_3^{*}$ and ${\boldsymbol{\Delta}}_{4,2}^{*}$ could help us better understand the evolution of constitutive relationships.

In practical applications, if a same-order kinetic moment of $f$ needs to be accurately known, it may be necessary to add the appropriate kinetic moment(s) of $f^{(0)}$ to be preserved according to the dependency given by CE. 
For example, in DBM considering up to the second-order TNE effect, if we need further to accurately know $\boldsymbol{\Delta}_{5,3}^{*}$ or $\bf{M}_{5,3}^{*}(f)$, then according to Eq.~(\ref{eq:B11}), we need only to add $\bf{M}_{6,4}^{*}(f^{(0)})$ to the list of kinetic moments to be preserved.

Here, the ${\boldsymbol{\Delta}} _3^*$ can be regarded as the flux of NOMF, and the ${\boldsymbol{\Delta}} _{4,2}^*$ can be regarded as the flux of NOEF, respectively. 
The first role of ${\boldsymbol{\Delta}} _{3}^*$ and ${\boldsymbol{\Delta}} _{4,2}^*$ is to help  understanding the evolution of stress and heat flux from a more fundamental level, and the second role is to help assessing the necessity of introducing second-order TNE effects in the constitutive relations of KMM. \cite{xu2022complex,zhang2022discrete}

It is obvious that \emph{roughly equivalent physical function to DBM are the extended hydrodynamic equations obtained by KMM.} In addition to the conservation equations of mass, momentum and energy, the extended hydrodynamic equations also contain the evolution equations describing some higher order kinetic moments such as ${{\bf{M}}_3^{*}}$ and ${{\bf{M}}_{4,2}^{*}}$.
The first function of ${\boldsymbol{\Delta}} _{n+1}^*$ is to help  understanding the evolution of ${\boldsymbol{\Delta}} _{n}^*$ from a more fundamental level, and the second function is to help assessing the necessity of introducing the $(n+1)$-th order TNE effects in the constitutive relations of KMM. \cite{xu2022complex,zhang2022discrete}

\nocite{*}

%

\end{document}